\documentclass[fleqn,usenatbib]{mnras}

\usepackage{newtxtext,newtxmath}

\usepackage[T1]{fontenc}

\DeclareRobustCommand{\VAN}[3]{#2}
\let\VANthebibliography\thebibliography
\def\thebibliography{\DeclareRobustCommand{\VAN}[3]{##3}\VANthebibliography}

\usepackage{multirow}
\usepackage{hyperref}
\usepackage{longtable}
\usepackage{pdflscape}
\usepackage{subcaption}

\usepackage{graphicx}	
\usepackage{amsmath}	

\newcommand {\kms} {\,{\rm km\,s}^{-1}}
\newcommand {\kpc} {\,{\rm kpc}}

\usepackage{xcolor}

\title[Fast rotations in galaxies at cosmic noon]{Fast rotations in galaxies at cosmic noon indicate central concentration of stars, dark matter or massive black holes}
\author[F. Roman-Oliveira et al.]{
Fernanda Roman-Oliveira,$^{1}$
Francesca Rizzo,$^{1}$\thanks{E-mail: rizzo@astro.rug.nl},
Filippo Fraternali$^{1}$
\\
$^{1}$Kapteyn Astronomical Institute, University of Groningen, Landleven 12, 9747 AD, Groningen, The Netherlands
}

\date{Accepted XXX. Received YYY; in original form ZZZ}

\pubyear{\the\year{}}

\begin{document}
\label{firstpage}
\pagerange{\pageref{firstpage}--\pageref{lastpage}}
\maketitle

\begin{abstract}
The rotation curves of regularly rotating disc galaxies are a unique probe of the gravitational potential and dark matter distribution. 
Until recently, matter decomposition of rotation curves at $z>0.5$ was challenging, not only due to the lack of high resolution kinematic data but also of both suitable photometry to accurately trace the stellar surface density and spatially-resolved sub-mm observations to trace the cold gas distribution.
In this paper, we analyse three galaxies from the Archival Large Program to Advance Kinematic Analysis (ALPAKA) sample, combining highly resolved cold gas observations from ALMA with rest-frame near-infrared imaging from JWST to investigate their dynamical properties and constrain their dark matter halos. 
The galaxies, initially classified as regularly rotating discs based on ALMA observations alone, appear in JWST as extended and symmetric stellar discs with spiral arms. 
Our dynamical models reproduce the rotation of the discs in the outer parts well, but they systematically underpredict the inner rotation velocities, revealing a deficit of central mass relative to the data. 
This discrepancy indicates either an underestimation of the bulge masses due to variations in the stellar mass-to-light ratio or dust attenuation or the presence of overmassive black holes. 
Alternatively, it may suggest departures from standard dark-matter halo profiles, including enhanced central concentrations. 

\end{abstract}


%

\section{Introduction} \label{sec5:intro}

Gaseous and stellar discs are a predominant morphological feature in local galaxies. Due to their regular structure and kinematics, they allow us to obtain information about their mass distribution and different mass components \citep{bosma78, rubin80, vanalbada85, mcgaugh04, delgado-serrano10, lelli16b}.
This provides a unique opportunity to study the interplay between baryonic and dark matter. In the local Universe, studies of neutral atomic hydrogen (HI) are the best for tracing the rotation curves of galaxies due to the radial extent of the HI emission and the quality of the interferometric data. These studies led the way to detailed mass models of rotation curves of hundreds of galaxies in the local Universe, ranging from dwarf galaxies to giant spirals \citep{lelli16, mancerapina22a, mancerapina22b, biswas23}.
This is possible due to the ability of a cold gas component, such as HI, to trace the circular speed and, effectively, the gravitational potential as a function of galactocentric radius.
Observational constraints on the mass distribution of the baryonic components, namely gas and stars, allow then to infer the properties of the dark matter halo \citep{deblok10, penarrubia10, tulin18, read19, posti19, posti21, nadler23, banareshernanez23, mancera2025}. Some of these studies have also opened debates over the shape of dark matter halos, such as the well known cusp-core problem, the nature of dark matter itself, and the impact of feedback on the stellar-to-halo mass ratios \citep{tulin18, read19, mancerapina22a}.

Deriving the properties of dark matter halos at $z >$ 0.5 through the analysis of rotation curves is significantly more challenging than at $z = 0$. The first difficulty is obtaining accurate kinematics, as the HI emission is no longer available due to the relatively low sensitivity of current radio telescopes \citep{verheijen07, jaffe12, fernandez16, chowdhury22}.
At $z \gtrsim 0.5$, kinematic studies of galaxies have relied on alternative tracers, mostly emission lines that trace warm ionised gas, such as H$\alpha$, [OII] and [OIII] \citep{wisnioski15, turner17, forster18, puglisi23, Danhaive2025}.
These emission lines are very bright and, therefore, they are the best tracer for obtaining large samples of hundreds of galaxies. 
However, the gravitational potential may not be optimally characterised, as the motions of the warm gas might be influenced by non-circular motions driven by Active Galactic Nuclei (AGN) or stellar feedback \citep[e.g.,][]{levy18, girard21, rizzo24, kohandel24, Phillips_2025}.

The second challenge concerns the information on the baryonic components, which is crucial for performing rotation curve decompositions.
Before the launch of \textit{JWST}, spatially resolved rest-frame near infrared observations of the stellar emission at $z >$ 1 were unavailable, leading to poor constraints on the stellar contribution to the total gravitational potential. 
Moreover, the bulk of the gas mass, which is expected to be in the form of cold atomic and molecular gas, is difficult to estimate due to the long integration time needed for observing CO or [CI].

As a consequence of the above, in previous attempts, the total gas mass has typically been fixed based on average values of the gas fraction expected for the star-forming population or estimated from scaling relations that correlate the molecular gas mass to the star formation rate \citep[e.g.,][]{tacconi18}. 
Additionally, the spatially-resolved gas mass profile has been assumed to follow the distribution of the warm gas \citep[e.g.,][]{genzel13, uebler18, genzel20, puglisi23, Ciocan_2025, Sharma_2025}.

A further complication arises when either the data quality (e.g., limited spatial or spectral resolution) or the intrinsic nature of the observations (e.g., slitless spectroscopy) prevents a reliable assessment of whether a galaxy hosts a disk whose gas kinematics is dominated by circular motions \citep{degraaff24, Lee_2025, Danhaive_2025b}. As a result, the inferred parameters of the dark-matter halo can become highly uncertain or even unreliable.

The above assumptions introduce several uncertainties, and an ideal way to overcome them is to combine high spectral and angular resolution observations of emission lines tracing cold gas from ALMA with rest-frame near-infrared observations from JWST. However, at $z = 1 – 3$, the only way to obtain spatially resolved kinematics of cold gas is to target CO or [CI] transitions. These lines are fainter than [CII] and, therefore, spatially-resolved observations of cold gas are available only for a handful of galaxies in this redshift range \citep[e.g.][]{noble19, molina19, kaasinen20, xiao22, lelli23}.

In a recent effort to address the lack of high-quality emission line observations of cosmic noon galaxies, the ALMA Archival Large Programme to Advance Kinematic Analysis (ALMA-ALPAKA) project was presented in \citet{rizzo23}. 
This is the most comprehensive sample of resolved cold gas observations in galaxies at $0.5 < z < 3.5$ and it attempts to bridge the gap in observational cold gas kinematic studies using medium to high-resolution ($\approx$ 0.2") observations of CO or [CI] emission lines from the ALMA archive.

The ALPAKA sample comprises 28 objects accounting for a total of $\approx 90$ hours of on target observation time and a total integration time of 147 hours, including overheads. In this paper, we focus on three well-resolved rotating disc galaxies that are part of the ALPAKA sample. 
These galaxies are chosen for the availability of resolved stellar continuum data obtained with JWST. 
We model the mass contribution of baryonic matter to the rotation curves to investigate the properties of their dark matter halos.
To our knowledge, this is the first mass decomposition using robust constraints on the baryons, with resolved rest-frame near-infrared JWST observations for the stars and cold gas kinematics from ALMA observations at $z >$ 0.5.

The paper is structured as follows: in Section~\ref{sec5:data}, we describe the data used; in Section~\ref{sec5:methods}, we describe the methodology employed in the analysis of the rotation curves; in Section~\ref{sec5:results}, we present our main results regarding the mass decomposition of the rotation curves; in Section~\ref{sec5:discussion}, we discuss our main findings; in Section~\ref{sec5:conclusions}, we summarise our results and their implications within the field. Throughout this work, we adopt a $\Lambda$CDM cosmology from the 2018 Planck results \citep{planck20}. 

\section{Data}\label{sec5:data}

\subsection{ALMA-ALPAKA subsample}

The ALPAKA project was designed to fill the gap of resolved cold gas kinematics at intermediate redshift (0.5 $<$ z $<$ 3.5) using public archival ALMA data of galaxies observed in CO or [CI]. 
This opens the opportunity to study the dark matter halos of cosmic noon galaxies in detail. However, to perform the mass modelling of the rotation curves, we need to first select suitable galaxies.

In \citet{rizzo23}, the 28 galaxies in the ALPAKA sample were classified according to their kinematics, and it was found that 19 galaxies are regularly rotating discs. The remaining galaxies were classified as merging systems (2/28) and uncertain (7). The latter means that the quality of the data is not good enough to robustly classify the galaxies as a disc or an interacting system.
For the present work, we selected only galaxies that: i) were classified as regular discs; ii) have at least three resolution independent elements over the major axis; iii) have public data from \textit{JWST}/NIRCAM; (iv) are at $z < 3$.
We note that, despite some galaxies having Hubble Space Telescope (\textit{HST}) imaging and/or an estimate of the stellar mass through Spectral Energy Distribution (SED) fitting, it does not necessarily mean that there is a good constraint on the stellar distribution. This is because the reddest \textit{HST} band covers, at most, the rest-frame optical and not the near-infrared.

Moreover, for relatively dusty galaxies, the light distribution is significantly attenuated, and we cannot robustly recover the stellar profile \citep{polletta24}. Therefore, to reduce this bias, our final subsample contains the three galaxies for which there is coverage of the rest-frame near-infrared. 
This emission comes from the bulk of the stellar populations and it is typically less affected by dust attenuation.
In Table~\ref{tab:data}, we list the ID and redshift of the ALPAKA subsample used in this work.

\begin{table*}
    \caption{Main properties and data of the three ALPAKA discs studied in this paper. The galaxies were selected based on their kinematic properties, spatial resolution and availability of ancillary \textit{JWST} data, as explained in the main text. Columns: (i) ALPAKA ID number. (ii) Redshift of the galaxy. (iii) Emission line used for the kinematics. (iv) Corresponding emission-line luminosity. (v) Number of independent resolution elements over the major axis of the ALMA data, defined as the diameter of the gas disc divided by the circularised ALMA beam. (vi) and (vii) \textit{JWST} imaging filter used for tracing the distribution of the stellar continuum and the corresponding rest-frame pivot wavelength. (viii) and (ix) Stellar mass and star formation rate, respectively \citep[see][for details]{rizzo23}.}
    \label{tab:data}
    \centering
    \begin{tabular}{cccccccccc}
    \hline\hline \noalign{\vskip 1mm}
    \multirow{2}{*}{ID} & \multirow{2}{*}{Redshift} & Emission line & \multirow{2}{*}{$ L' \left( \frac{10^{10} \mathrm{K} \hspace{1mm} \mathrm{km} \hspace{1mm} \mathrm{pc}^2}{\mathrm{s}} \right)$ } & \multirow{2}{*}{\# RE}  & \textit{JWST} filter &  Rest-frame $\lambda$ & \multirow{2}{*}{$\log \left( \frac{M_{*}}{M_\odot} \right)$} & SFR   \\ 
    & & &  & &  & ($\mu$m) & & ($M_\odot$ yr$^{-1}$) & \\  \hline \noalign{\vskip 1mm} \vspace{2mm}
    1	& 0.561		& CO~(2-1)	    & 0.14 $\pm$ 0.03 & 3.8	& F277W  & 1.78 & 10.04$^{+0.10}_{-0.14}$ & 8 $\pm$ 2  \\ \vspace{2mm}
    3	& 1.445		& CO~(5-4)	    & 1.5 $\pm$ 0.3 & 4.2	& F444W  & 1.81 & 10.51$^{+0.13}_{-0.19}$ & 281 $\pm$ 18 \\
    13	& 2.103		& [CI]~($^3$P$_2$ - $^3$P$_1$)	& 0.90 $\pm$ 0.33 & 4.2	& F444w  & 1.42 & 11.08$^{+0.10}_{-0.11}$ & 230 $\pm$ 56  \\
    \noalign{\vskip 1mm}
    \hline
    \end{tabular}   
\end{table*}

\subsection{ALMA data}

The ALMA datacubes used in this work were calibrated and imaged as described in \citet{rizzo23}.
For this work, we collapsed the continuum subtracted datacubes in the velocity direction to create moment 0 maps of the CO/[CI] emission line. The velocity range of the original cubes extends for at least $\approx 800 \kms$. These maps are centred in the galaxy and have wide enough fields of view to encompass the whole emission. We used these maps to model the gas surface brightness profile. 
We also used the rotation velocity profiles as a function of radius, obtained through the kinematic models with \texttt{$^{\mathrm{3D}}$BAROLO}, reported in \citet{rizzo23}. In Figure~\ref{fig:data}, we overlay the gas moment 0 maps on top of the rest-frame near-infrared data of our sample (discussed below).

\subsection{Rest-frame optical and near-infrared data}

\begin{figure*}
    \centering
    \includegraphics[width=0.95\textwidth]{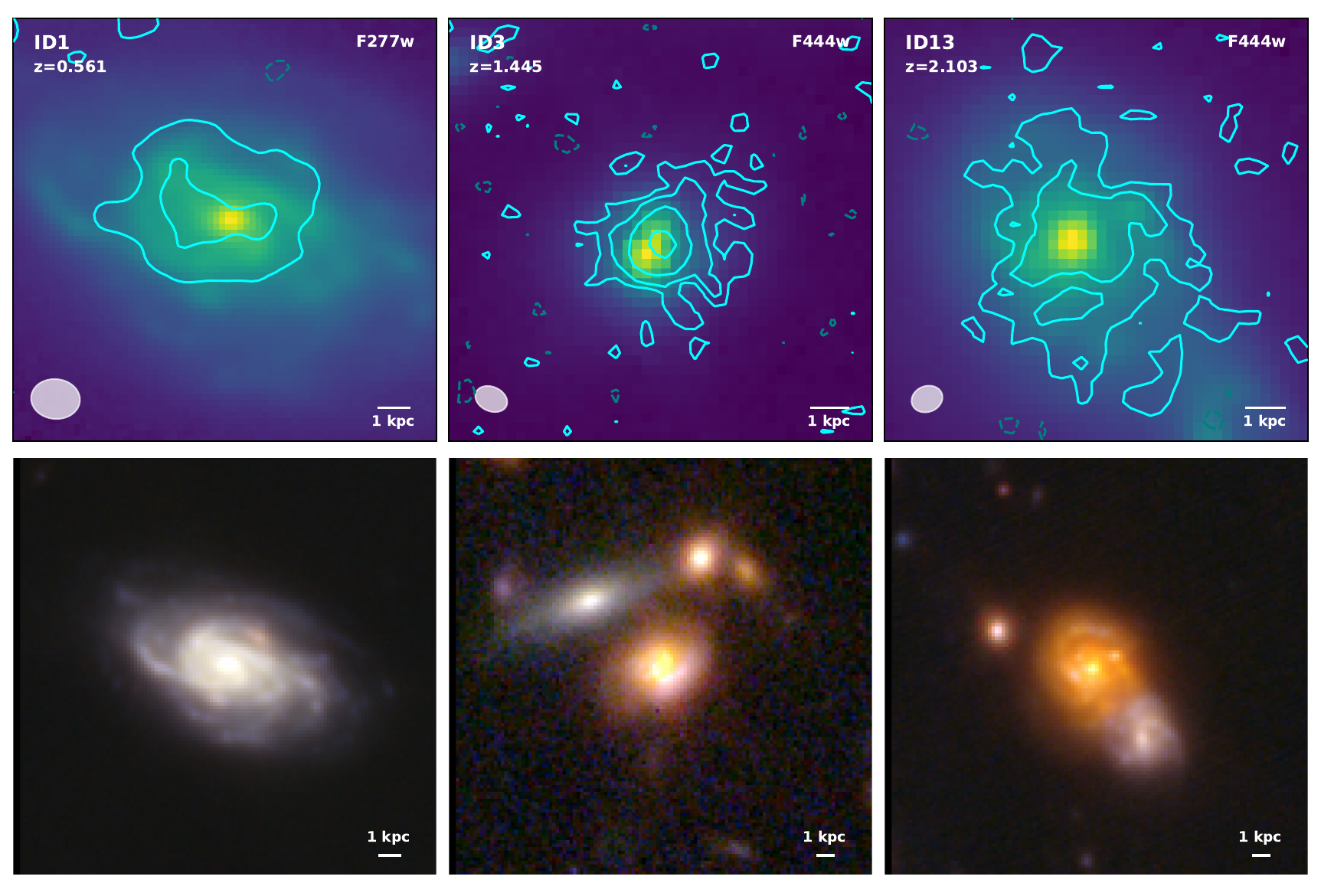}
    \caption[\textit{JWST} and ALMA images for selected ALPAKA galaxies.]{Top row: ALPAKA subsample of galaxies analysed in this work. The galaxies are organised in ascending order of redshift, from left to right. The gas data (see Table~\ref{tab:data}) are shown as blue contours with levels following $3 \sigma$, $6 \sigma$ and $12 \sigma$, where $\sigma$ is the root-mean-square (RMS) noise of the ALMA moment 0th map of the gas emission. We also show negative emission with dashed contours at the $-3\sigma$ level and the corresponding ALMA beam on the bottom left of each panel. In the background, we display the \textit{JWST} filter used for deriving the stellar surface brightness, same as the one reported in Table~\ref{tab:data}. 
    The ALMA contours shown here are simply the emission levels on the unmasked 0th-moment maps, therefore they differ from the pseudo-X $\sigma$ contours shown in \citet{rizzo23}. Bottom row: RGB composite \textit{JWST} images. We note that the scales in the top and bottom rows are different with a larger FOV on the bottom panels that allow us to see the foreground and background galaxies surrounding ID3 and ID13.} 
    \label{fig:data}
\end{figure*}

For our three targets, we use publicly available \textit{JWST}/NIRCam data from NIRCam, processed as part of the DAWN JWST Archive \citep[DJA,][]{Valentino_2023, Heintz_2025}.
To have homogeneous coverage in the analysis of the stellar continuum distribution, we select the \textit{JWST} filters that cover the rest-frame emission at 1.5 -- 2 $\mu$m, corresponding to F277W for ID1 and F444W for ID3 and I13 (Table~\ref{tab:data}).
In Figure~\ref{fig:data}, we show a colour-composited image using the \textit{HST} and \textit{JWST} images and in Table~\ref{tab:data} we report the rest-frame wavelengths of the filters used.

\section{Methods}\label{sec5:methods}

\subsection{2D Surface Brightness Fitting}\label{sec5:dysb}
To estimate the surface density profile used for the rotation curve decomposition, we derive the structural parameters of both the gas and the stellar distributions. The stellar distributions of these three galaxies are, in general, more complex than the gas distributions due to the presence of other galaxies in the foreground/background. 
For this reason, we used \texttt{GALFIT} to derive models of the stellar component, including disc and bulge components, and we used our own tool, as described below, to estimate the best-fit parameters of the gas distribution with a single disc component.
For the stars, the \textit{JWST}/F277w imaging for ID1 and \textit{JWST}/F444w imaging for ID3 and ID13 correspond to rest-frames of 1.78, 1.81 and 1.42 $\mu$m respectively (see columns v and vi of Table~\ref{tab:data}).
For the gas, we used the moment 0 maps of the CO or [CI] emission line transitions available from the ALPAKA survey (see the third column in Table~\ref{tab:data}).

\subsubsection{Gas distribution} \label{sec:gasdis}

To fit the gas distribution of the three ALPAKA targets, we used the methodology described in \citet{romanoliveira24} assuming that they follow an exponential profile, corresponding to a Sérsic profile \citep{sersic68} with the Sérsic index $n$ fixed at 1. In addition to the parameters defining the profile (i.e., intensity at the effective radius $I_{\mathrm{eff}}$, effective radius $R_{\mathrm{eff}}$), we fitted the geometrical parameters: centre, defined by $x_0$ and $y_0$, ellipticity $\epsilon$, and position angle $PA$.
We fitted the 2D surface brightness model using \texttt{DYNESTY}, a python package for estimating Bayesian posteriors and evidences with the Dynamic Nested Sampling algorithm \citep{skilling04, skilling06, speagle20, koposov22}. Before fitting the data, we convolve the model to match the same spatial resolution as the data. 
Moreover, similarly to \citet{romanoliveira24} and \citet{marasco19}, we modified the minimisation function to better represent the errors in the data. We used a chi-square likelihood with residuals weighted by an effective uncertainty that depends on the asymmetry of the galaxy, such that: $(|\mathcal{D} - \mathcal{M}|)^2 / \sigma_{\mathrm{eff}}^2$, where $\mathcal{D}$ is the data and $\mathcal{M}$ is the model. The effective uncertainty ($\sigma_{\mathrm{eff}}$) is defined as the standard deviation of the residual map obtained by rotating the data map by 180$^\circ$ and subtracting it from the original, non-rotated data map.
This provides a more realistic (and larger) estimate of the errors in the best-fit parameters, as the source of mismatch in the fitting is primarily due to deviation from the axisymmetry assumed in the model rather than the background noise in the data. In Table~\ref{tab:priorsdysb}, we summarise the priors used in the fitting of the gas surface brightness.

\begin{table}
    \caption{Priors for the parameters defining the 2D gas surface brightness models: intensity at the effective radius ($I_{\mathrm{eff}}$), effective radius ($R_{\mathrm{eff}}$), geometric centre ($x_0$, $y_0$), ellipticity ($\epsilon$) and position angle ($PA$). For the intensity, we arbitrarily chose 1.2 $\sigma_{\mathrm{RMS}}$ as the lower bound of the prior. 
    $\sigma_{\mathrm{RMS}}$ is the RMS noise in the map. 
    }
    \label{tab:priorsdysb}
    \centering
        \begin{tabular}{cccc}
        \hline\hline \noalign{\vskip 1mm}
        Parameter   		& Prior Type  & Value Range	 & Units \\ \hline \noalign{\vskip 1mm}
        $I_{\mathrm{eff}}$               & Uniform   & 1.2$\sigma_{\mathrm{RMS}}$ -- max in map &  Jy beam$^{-1}$ $\kms$	\\
        $R_{\mathrm{eff}}$  & Uniform   & $\sim$ 1.3 -- $\sim$ 13 & kpc	\\
        $x_0$                & Uniform   & image centre $\pm$ 5 & kpc	\\ 
        $y_0$                & Uniform   & image centre $\pm$ 5 & kpc 	\\ 
        $\epsilon$          & Uniform   & 0.01 -- 1.0 & - \\ 
        PA                  & Uniform   & 0 -- 360 & degrees 	\\ 
        \hline
        \end{tabular}

\end{table}

\subsubsection{Stellar distribution}
The code used to model the gas distribution (Sect.~\ref{sec:gasdis}) was originally developed for ALMA data, where the beam is well approximated by a Gaussian kernel. In contrast, the \textit{JWST} point spread function (PSF) is considerably more complex and cannot be accurately described by a Gaussian profile. Because adapting and running the gas-distribution code with the full \textit{JWST} PSF would be computationally expensive, we instead employed \texttt{GALFIT} \citep{Peng_2010} to model the stellar surface brightness. To characterise the \textit{JWST} PSF, we identified an unsaturated star within 30'' of each target galaxy. As shown in Figure~\ref{fig:data}, for ID3 and ID13 the modelling also required simultaneous fits to nearby foreground and background sources in order to obtain reliable structural parameters for the main galaxies.

For the three targets in our sample, we fitted the main galaxy with two components (one for the bulge and another for the disc) and an additional flat component for the sky background. For ID3 and ID13, we also fit extra components for the foreground and background galaxies, for which we assumed a S\'ersic profile \citep{sersic68} and let the total intensity, effective radius and S\'ersic index free to vary.

For the disc components of the target galaxies, we assumed an exponential profile. For each galaxy, we first  fit the disc alone, thus determining its total intensity, effective radius, centre, ellipticity, and position angle. 
In the subsequent run, which included the bulge component, we fixed the geometrical parameters (centre, ellipticity, and position angle) in order to reduce the number of free parameters.
For the bulges, we used a de Vaucouleurs profile (i.e., S\'ersic index with $n = 4$) and fixed their centres to coincide with the disc centre.
Given that, in our test runs, the bulges were about the size of the PSF in all three cases, we fixed the bulge effective radius to half the FWHM of the PSF.
We also fix the flux of the bulge to values that maximise their brightness within the uncertainties in the data.
We do this for our fiducial models because, as we show in more detail in Section~\ref{sec5:res_dysb} (see also Appendix \ref{sec5:dysb_corner}), the rotation curves indicate the presence of a central compact component.
However, given that the bulge contributes little to the global emission and has a size comparable to the PSF, the \texttt{GALFIT} fit will always tend to underestimate its brightness.
This maximisation of the bulge components is a conservative choice against our main results, as we see below.

\subsection{Circular Speed}\label{met:adc}

Assuming that the gravitational potentials ($\Phi$) of the ALPAKA galaxies are axisymmetric, the circular speed ($V_{\mathrm{c}}$) of a test particle moving in perfect circular orbits in the midplane under the influence of $\Phi$ is given by
\begin{equation}
    V_{\mathrm{c}}^2 = R \: \left( \frac{\partial \Phi}{\partial R} \right),
\end{equation}
\noindent
where $R$ is the radius of each orbit.
We can infer the radial profile of the circular speed of the galaxies by applying the asymmetric drift correction to the measured rotation of the gas. This correction accounts for pressure support and can be described by
\begin{equation}\label{eq5:vc}
    V_{\mathrm{c}}^2 = V_{\mathrm{rot}}^{2} + V_{\mathrm{A}}^{2} = V_{\mathrm{rot}}^{2} - R \sigma^2 \frac{\partial \ln (\sigma^2 \Sigma_{\rm gas} )}{\partial R},
\end{equation}
\noindent
where $V_{\mathrm{A}}$ is the correction for asymmetric drift, $V_{\mathrm{rot}}$ and $\sigma$ are the rotation velocity and velocity dispersion of the gas obtained from kinematic modelling with \texttt{$^{\mathrm{3D}}$BAROLO}, and $\Sigma_{\mathrm{gas}}$ is the surface density of the gas. 
Here, we assume that the disc thickness is radially constant. 
To see this derivation more explicitly, we refer to \citet{iorio17}.
We note that in galaxies dominated by rotational motion ($V/\sigma \gg 1$), this correction is usually minimal. 
However, different regions of the same galaxy may have different pressure support and will be affected differently as the correction is applied to each point in the radial profile. 
This can have an impact on the overall shape of the rotation curve.

Assuming that the gas is distributed as an exponential disc, we can rewrite the asymmetric drift term $V_{\mathrm{A}}$ in Equation~(\ref{eq5:vc}) as
\begin{equation}
V_{\mathrm{A}}^2 = - R \sigma^2 \frac{\partial}{\partial R} \left[\ln{(\sigma^2  \exp{(-R/R_{\mathrm{d, gas}}}})) \right],
\end{equation}
\noindent
and further,
\begin{equation}
V_{\mathrm{A}}^2 = - 2 R \sigma^2 \frac{\partial \ln \sigma}{\partial R}  + \frac{R \sigma^2}{R_{\mathrm{d, gas}}},
\end{equation}
\noindent
where $R_{\mathrm{d, gas}}$ is the disc-scale length of the exponential gas disc that is defined as $R_{\mathrm{eff}}/1.678$. To apply this correction, we use $\sigma$ and $R$ from the kinematic values derived in \citet{rizzo23}, we then need to derive $R_{\mathrm{d, gas}}$ for each galaxy as we explain in Section~\ref{sec5:dysb}.

\subsection{Rotation Curve Decomposition}

To model the circular speed of the ALPAKA galaxies, we assume a model circular velocity as defined by 
\begin{equation}\label{eq:vall}
    V_{\mathrm{c, tot}}^2 = \sum_{i=1}^n R \left(\frac{\partial \Phi_{\mathrm{i}}}{\partial R} \right)  =  \sum_{i=1}^n V^2_{\mathrm{c, i}} \hspace{1mm},
\end{equation}
where $\Phi_i$ and $V_i$ are the gravitational potential and circular velocity of the i-th mass component (i.e., stellar bulge, stellar disc, gas disc, and dark matter), modelled as described in the following:

\begin{itemize}
    \item Stellar Bulge: The projected light distribution is assumed to follow a de Vaucouleurs profile. To describe the circular velocity contribution of the bulge ($V_{\mathrm{c, b}}$), we adopt the generalised deprojected spherical S\'ersic profile of \citet{terzic05}, which gives
\begin{equation}\label{eq:vsersic}
    V_{\mathrm{c, b}}^2 = \frac{G M_{\mathrm{b, *}}}{R} \frac{\gamma(n(3-p)),b(R/R_{\mathrm{eff},*})^{1/n}}{\Gamma(n(3-p))},
\end{equation}
where $M_{\mathrm{b, *}}$ is the stellar mass of the bulge, $R_{\mathrm{eff},*}$ is the bulge effective radius, and $n$ is the S\'ersic index, which we fix to 4.0, equivalent to a de Vaucouleurs profile. 
The parameters $p$ and $b$ are functions of $n$, and $\gamma$ and $\Gamma$ are the incomplete and complete gamma functions, respectively. 

\item Stellar Disc: The stellar disc is described by a thin exponential disc \citep{binney08}, whose circular speed is given by  
\begin{equation}\label{eq:expdisc}
    V_{\mathrm{c, d}}^2 = 4 \pi G \Sigma_0 R_{\mathrm{d, *}} y^2 (I_{0}(y)K_{0}(y) - I_{1}(y)K_{1}(y)),
\end{equation}
where $R_{\mathrm{d, *}}$ is the scale length of the stellar disc, $I_{0}, I_{1}, K_{0}, K_{1}$ are modified Bessel functions, y = $R/(2 R_{\mathrm{d, *}})$ and $\Sigma_0$ is the central surface density of the stellar disc. 
Stellar masses ($M_{\mathrm{*}}$) were derived from SED fitting (Table~\ref{tab:data}), corrected for aperture, i.e. corresponding to the total mass of an infinitely extended exponential profile. The central surface density is therefore  
\begin{equation}
    \Sigma_{0,*} = \frac{M_{\mathrm{d, *}}}{2 \pi R_{\mathrm{d, *}}^2},
\end{equation}
where $M_{\mathrm{d, *}}$ is the total mass of the stellar disc.

\item Gas Disc: The gaseous disc is also modelled as a thin exponential disc (Eq.~\ref{eq:expdisc}), but truncated at the observed CO/[CI] radius since no aperture correction is applied to molecular gas masses. Following \citet{binney08}, the central surface density is  
\begin{equation}
    \Sigma_{0,\rm gas} = \frac{M_{\mathrm{gas}}} {2 \pi R_{\mathrm{d, gas}}^2  \left[1 - \exp{\left(-\frac{R_t}{R_{\mathrm{d, gas}}}\right)} \left(1 + \frac{R_t}{R_{\mathrm{d, gas}}}\right)\right]},
\end{equation}
where $M_{\mathrm{gas}}$ is the total gas mass present in the disc and the truncation radius $(R_{\mathrm{t}})$ is defined as the extent of the gas disc.

\item Dark Matter Halo: For the dark matter distribution, we assume a spherical Navarro–Frenk–White (NFW) halo profile \citep{navarro97}. The circular speed is given by
\begin{equation}
\begin{split}
V_{\mathrm{DM}}^2 = \frac{G M_{200}}{R_{200}} \frac{\ln{(1+ c_{200} x)} - c_{200} x / (1 + c_{200} x)}{x[\ln{(1+c_{200})}- c_{200} / (1+c_{200})]},
\end{split}
\end{equation}
where $c_{200}$, $M_{200}$ and $R_{200}$ are the concentration of the dark matter halo, the virial mass and the virial radius, respectively, and $x = R/ R_{200}$.
\end{itemize}

Finally, we also explored the possibility of the contribution of a central supermassive black hole (SMBH) to the total circular speed of the galaxies. We assume that the circular speed of the SMBH ($V_{\rm c, \bullet}$) is described by a central point mass following
\begin{equation}
    V^2_{\rm c, \bullet} = \frac{M_{\bullet} G }{R},
\end{equation}
where M$_\bullet$ is the black hole mass.

\begin{table*}
    \caption{Rotation curve decomposition priors: fiducial model. The main three parameters we fit are the total gas mass, the total stellar mass and the dark matter mass. 
    The total gas mass is computed by the product of the CO/[CI] luminosity (see Table~\ref{tab:data}) and the gas normalisation, i.e. the CO/[CI] line ratio ($r_{l}$) times the $\alpha_{\mathrm{CO}/\mathrm{[CI]}}$ conversion. 
    The truncation radius is assumed to be the extent of the gas data, that is the last radius in the CO/[CI] rotation curve. The total stellar mass is let free to vary and it is computed by the sum of the contribution of the disc and bulge masses assuming that the mass-to-light ratio of the bulge is 1.4 times that of the disc. 
    The dark matter mass is computed from the baryonic fraction parameter that is let free to vary uniformly in log space.}
    \label{tab:fiducialprior}
    \centering
        \begin{tabular}{ccccc}
        \hline\hline \noalign{\vskip 1mm}
        Mass Component & Parameter  & Prior Type & Value Range & Units \\ \noalign{\vskip 1mm} \hline

\noalign{\vskip 1.5mm}
        \multirow{2}{*}{Dark matter} & Baryonic fraction, $f_{\mathrm{bar}}$ (effectively $M_{200}$)  & Uniform log & $-7$ : $-0.72$    & -\\
        & NFW concentration, $c_{200}$  & Fixed      & CMR \citep{dutton14}  & -  \\ \noalign{\vskip 1.6mm} 

\noalign{\vskip 1.5mm}
        \multirow{4}{*}{Gas Disc}  & Gas mass ($M_{\rm gas}$) normalisation: $\alpha_{\mathrm{CO}/\mathrm{[CI]}} \times r_{l}$  & {Uniform} & {0.1 : 8} & {$10^{10}$ K km s$^{-1}$ pc$^2$}\\
        & Disc scale length, $R_{\mathrm{d,gas}}$  & Fixed & $R_{\mathrm{d,CO}}$ or $R_{\mathrm{d,[CI]}}$ & $\mathrm{kpc}$\\
        & Truncation radius, $R_{\mathrm{t}}$  & Fixed & $R_{\mathrm{ALMA, max}}$ & $\mathrm{kpc}$\\ \noalign{\vskip 1.6mm}

\noalign{\vskip 1.5mm}
        \multirow{3}{*}{Stars} & Total stellar mass, $M_{*}$   & Uniform log    & 9.5 : 13.0  & M$_{\odot}$ \\ 
        &  Disc effective radius, $R_{\mathrm{eff, disc}}$  & Fixed    & $R_{\mathrm{eff,JWST}}$ & $\mathrm{kpc}$ \\
         & Bulge effective radius, $R_{\mathrm{eff, b}}$  & Fixed  & \textit{JWST} PSF FHWM/2 & $\mathrm{kpc}$ \\ 
        \hline
        \end{tabular}
\end{table*}

\subsubsection{Modelling the rotation curves}\label{sec5:modelling}
We fitted the galaxy rotation curves, corrected for asymmetric drift, using Equation~(\ref{eq:vall}). The posterior distribution of the model parameters was explored using the nested-sampling algorithm \texttt{DYNESTY} \citep{speagle20}. The fits were evaluated by minimising the residuals through a chi-square likelihood, $(V_{\mathrm{c}} - V_{\mathrm{c, tot}})^2 / V_{\mathrm{err}}^2$, where $V_{\mathrm{err}}$ represents the uncertainty in the observed rotation curve.

In our fiducial models, we used the distribution of the gas and stellar surface density obtained by fitting the ALMA and \textit{JWST} data. 
We kept them fixed while leaving their normalisation (i.e., $M_{\mathrm{gas}}$, $M_* = M_{\mathrm{b,*}} + M_{\mathrm{d,*}}$) free to vary. 
We included the NFW halo with a concentration following the concentration mass relation (CMR) from \citet{dutton14}, which is described as
\begin{equation}
    \log_{10} c_{200} = a + b \log_{10} \left( \frac{M_{200}}{10^{12} h^{-1} M_{\odot}} \right), 
\end{equation}
where $h^{-1}$ is the reduced Hubble constant, $a$ and $b$ are described by
\begin{align}
        a & = 0.520 + (0.905 - 0.520) \exp{(-0.617 z^{1.21})},\\
    b & = -0.101 + 0.026 z,
\end{align}
and $z$ is the redshift of the source.

The mass of the dark matter halo ($M_{200}$) was calculated indirectly from a baryonic fraction parameter ($f_{\mathrm{bar}}$), defined as the ratio between the baryonic mass ($M_{\mathrm{bar}}$) and the dark matter halo virial mass, $f_{\mathrm{bar}}$ = $M_{\mathrm{bar}}/M_{200}$. The baryonic fraction parameter was kept free with a flat prior. 
The upper limit of the prior is given by the cosmological baryonic fraction of 18.7\% from \citet{planck20}, ensuring that there is a physically motivated minimal amount of dark matter contributing to the rotation curves within R$_{200}$. The lower limit was set to a baryonic fraction of 10$^{-7}$, which corresponds to a heavily dark matter dominated galaxy. This lower limit is motivated to set a wide enough prior range for the posteriors so that they are not affected by the edges of the prior, as the baryonic fraction tends to have a tail towards the lower end (see Figure~\ref{fig:dydy_ID1_corner}).

For the gas, we computed the total gas mass (M$_{\mathrm{gas}}$) according to
\begin{equation}\label{eq:alpha}
    \mathrm{M}_{\mathrm{gas}} = L_{\mathrm{CO/[CI]}} \hspace{1mm} r_{l} \hspace{1mm} \alpha_{\mathrm{CO/[CI]}}, 
\end{equation}
where $L_{\mathrm{CO/[CI]}}$ is the CO/[CI] luminosity, $r_l$ is the luminosity line ratio and $\alpha_{\mathrm{CO/[CI]}}$ is the CO/[CI] luminosity-to-H2 mass conversion.
We kept the product of $r_{l} \hspace{1mm} \times \alpha_{\mathrm{CO/[CI]}}$ as a free parameter in the fiducial models with a flat prior. 
In Section~\ref{sec5:dm}, instead we kept these parameters fixed as we attempted to constrain the dark matter component by fixing the baryonic components. 
In this case, we chose an $\alpha_{\mathrm{CO}}$ value of 3.6 and 0.8 M$_\odot /10^{10} \mathrm{K} \hspace{0.8mm} \mathrm{km} \hspace{0.8mm} \mathrm{s}^{-1}\mathrm{pc}^2$ for ID1 and ID3, as these are the typical values used for main-sequence and starburst galaxies, respectively \citep{daddi10a, kirkpatrick19}.
For ID13, we used the [CI]~($^3$P$_2$ - $^3$P$_1$) emission to derive the total molecular mass after assuming an $\alpha_{\mathrm{[CI]}}$ factor.
Using astrochemical simulations, \citet{bisbas21} suggested that the $\alpha_{\mathrm{[CI]~(^3P_1 - ^3P_0)}}$ values range from 1 to 9.7 M$_\odot /10^{10} \mathrm{K} \hspace{0.8mm} \mathrm{km} \hspace{0.8mm} \mathrm{s}^{-1}\mathrm{pc}^2$, with the most common value being 7 M$_\odot /10^{10} \mathrm{K} \hspace{0.8mm} \mathrm{km} \hspace{0.8mm} \mathrm{s}^{-1}\mathrm{pc}^2$. Other observational works on both local and high-$z$ galaxies have found values that roughly agree with this range \citep{valentino18, crocker19, valentino20b, heintz20, lelli23}. 
Another study focused on submillimitre galaxies has found somewhat higher values of $\sim 15$ M$_\odot /10^{10} \mathrm{K} \hspace{0.8mm} \mathrm{km} \hspace{0.8mm} \mathrm{s}^{-1}\mathrm{pc}^2$ \citet{dunne22}.
In Section~\ref{sec5:dm}, we chose a conservative $\alpha_{\mathrm{[CI]}}$ of 1 M$_\odot /10^{10} \mathrm{K} \hspace{0.8mm} \mathrm{km} \hspace{0.8mm} \mathrm{s}^{-1}\mathrm{pc}^2$, the lower end of the range of values reported in the literature. This is because larger values of $\alpha_{\mathrm{[CI]}}$, combined with the estimated stellar mass (Table~\ref{tab:data}) would overpredict the circular speed in the outer parts of the disc. 

Regarding the luminosity line ratios in Section~\ref{sec5:dm}, we used the results in \citet{boogaard20} for the CO emission, taking the values from their Table 3 for the subsamples at $1.0 < z < 1.6$ for ID1 and ID3. 
We note that this is the best collection of line ratios for galaxies in this redshift range, but are uncertain. 
It contains only 12 galaxies to calibrate the line ratio models, and only three of those were observed in CO~(5-4), the same transition as ID3.
For ID13, as discussed in \citet{valentino18}, the simple structure of neutral atomic carbon emission allows for the breaking of the temperature-density degeneracy that is not possible in the case of CO. 
Therefore, the excitation temperature of the gas can be directly related to the luminosity line ratio by T$_{\mathrm{ex}} = 38.8 \mathrm{K} / \ln{(2.11/r_l)}$ \citep{schneider03, weiss03}. 
Since we only have one [CI] transition available, we assumed a gas excitation temperature of $30\,{\rm K}$, as found on average by other works at a similar redshift \citep{walter11, valentino20b, henriquezbrocal22}. 
We obtained an estimate of the L'$_{\mathrm{[CI] ^3 P_2 - ^3 P _1}}$/L'$_{\mathrm{[CI] ^3 P_1 - ^3 P _0}}$ line ratio of 0.57.

For the stellar distribution, we adopted the structural parameters of the bulge and disc derived with \texttt{GALFIT}, keeping them fixed during the modelling. The bulge-to-disc luminosity ratio was also taken from the \texttt{GALFIT} decomposition. We assumed that the bulge mass-to-light ratio ($\Upsilon$) is 1.4 times higher than that of the disc, following the approach of \citet{lelli16b} for spiral galaxies at $z = 0$. As discussed in Section \ref{sec5:discussion}, the different mass-to-light ratios for the bulge and the disc components can be attributed to a difference in their stellar populations \citep[e.g.,][]{schombert14} or a gradient in dust attenuation.
In our fiducial fit (Section~\ref{sec5:fid}), we allowed the total stellar mass to vary independently of the prior stellar masses derived from SED fitting; however, we kept the relative contributions of mass in the disc and bulge fixed. 

In summary, for the fiducial models we have the following components with their corresponding parameters: (i) de Vaucouleurs stellar bulge with bulge mass and bulge effective radius ($M_{\mathrm{b, *}}, R_{\mathrm{eff,b}}$); (ii) stellar thin exponential disc, with stellar disc mass and stellar disc scale-length ($M_{\mathrm{d, *}}, R_{\mathrm{d,*}}$); (iii) gas truncated thin exponential disc, with gas mass, gas disc scale-length and truncation radius ($M_{\mathrm{gas}}, R_{\mathrm{d,gas}}, R_t$); and (iv) NFW halo, with dark matter mass and halo concentration ($M_{200}, c_{200}$). 
We summarise the priors used in the rotation curve decomposition of the three galaxies in Table~\ref{tab:fiducialprior}.

\begin{figure*}
    \centering
    \includegraphics[width=0.99\linewidth]{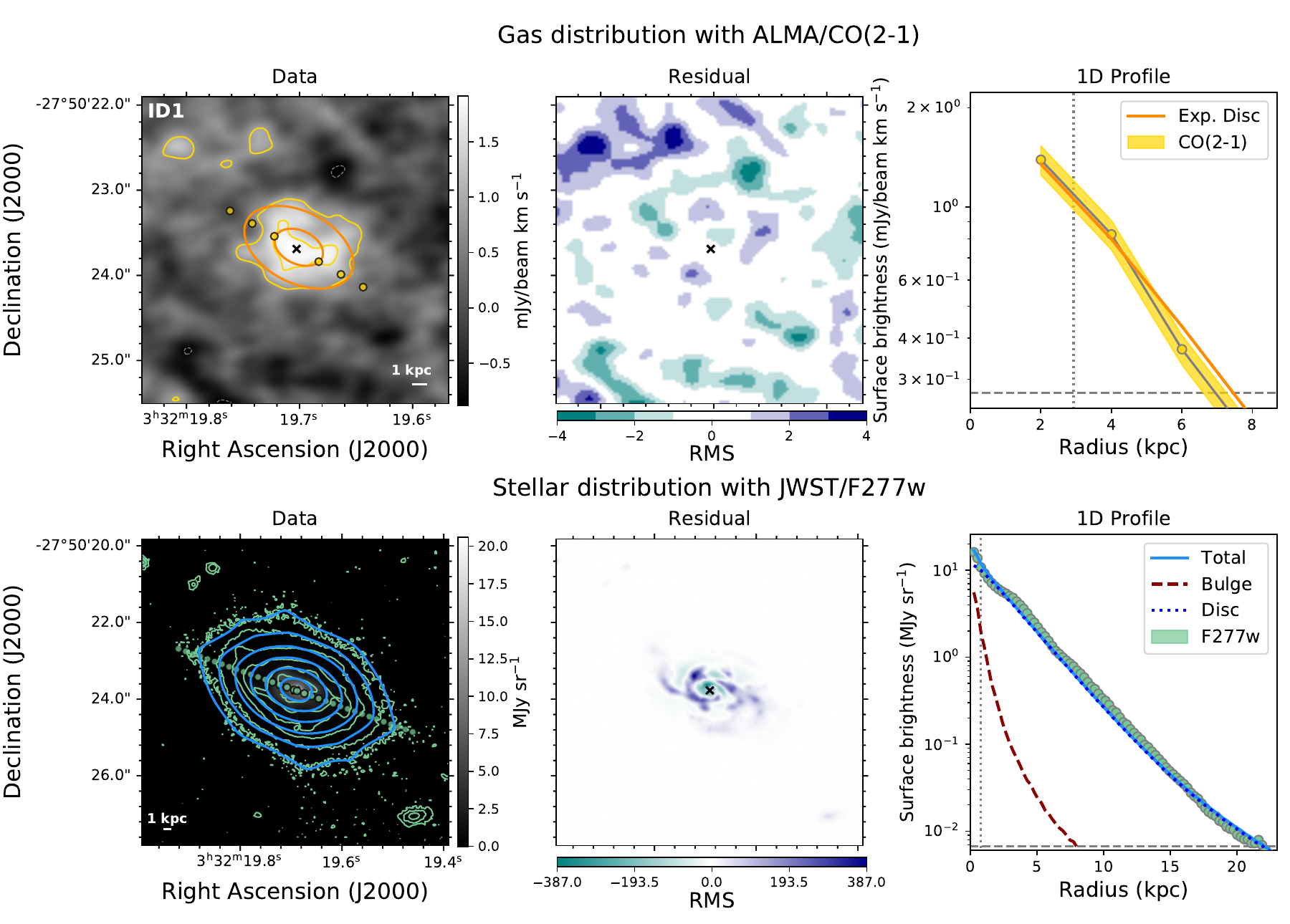}
    \caption[Surface brightness models of the gas and stellar distribution of ID1.]{Surface brightness models of the gas (top panels) and stellar distribution (bottom panels) of ID1. 
    Top left panel: the ALMA moment 0 map is overlayed with yellow contours for the data and orange for the best-fit model, following levels of 3$\sigma$, 6$\sigma$, 9$\sigma$, etc., where $\sigma$ is the RMS noise in the data. We also show negative contours with dashed gray lines at levels $-3\sigma$ and $-6\sigma$. 
    Top middle panel: noise-normalised residual map defined by (Data $-$ Model) / RMS. 
    Top right panel: 1D representation of the projected average surface brightness over inclined rings along the major axis of the galaxy. The yellow band represents the data and its uncertainties while the orange solid line represents the best-fit model. 
    Bottom left panel: \textit{JWST} image in grayscale is overlayed with teal contours for the data and blue for the best-fit model, following levels of 3$\sigma$, 9$\sigma$, 27$\sigma$, etc. above the background, where $\sigma$ is the RMS noise in the data. 
    Bottom middle panel: noise-normalised residual map defined by (Data $-$ Model) / RMS. 
    Bottom left panel: 1D representation of the projected average surface brightness over inclined rings along the major axis of the galaxy. The teal band represents the data and the blue solid line represents the total best-fit model which is composed by a stellar disc (dotted dark blue line) and stellar bulge (dashed red line). 
    In both top and bottom right panels we show the circularised ALMA beam/\textit{JWST} PSF and the data RMS as vertical dotted and horizontal dashed lines, respectively.}
    \label{fig:sb_id1}
\end{figure*}

\begin{table*}
\caption{Best-fit parameters of the 2D surface brightness fits to the gas and stars. The gas discs were modelled using our \texttt{DYNESTY} based script and the stellar discs and bulges were modelled using \texttt{GALFIT}. For the latter, we employed a two-iteration fitting process in which the geometric parameters (indicated with an asterisk) were fixed in the second iteration to reduce the number of free parameters. Both the gas and stellar disc are exponential discs and the bulge is a de Vaucouleurs profile. $R_{\rm eff}$ for the exponential disc is calculated as $1.678\times R_{\rm d}$}.
\label{tab:dysb}
\centering
\begin{tabular}{cc ccc ccc}
        \hline\hline \noalign{\vskip 1mm}
    & Component & $R_{\mathrm{eff}}$ & $x_0$ & $y_0$ & $\epsilon$ & $PA$ & B/T \\
    & & (kpc) & (ra) & (dec) &  & (deg) & \\\hline \noalign{\vskip 1.5mm}

 & Gas disc        & 5.0 $\pm$ 0.3 & 3:32:19.660 & -27:50:23.380 & 0.48 $\pm$ 0.03 & 60 $\pm$ 2  & -  \\ 
ID1 & Stellar disc   & 3.86 $\pm$ 0.01 & 3:32:19.693$^{*}$ & -27:50:23.795$^{*}$ &  0.38$^{*}$ &  73.3$^{*}$ & \multirow{2}{*}{0.05} \\ 
& Stellar bulge     & 0.4$^{*}$ &  " &  " &  -  &  - &  \\ \noalign{\vskip 1.7mm} 
 
 & Gas disc        & 1.74 $\pm$ 0.04 & 9:59:55.554 & 2:15:11.875 & 0.04 $\pm$ 0.02 & 117 $\pm$ 29  & -  \\ 
ID3 & Stellar disc   & 1.92 $\pm$ 0.02 & 9:59:55.552$^{*}$ & 2:15:11.481$^{*}$ &  0.14$^{*}$ &  134.1$^{*}$ & \multirow{2}{*}{0.2} \\ 
& Stellar bulge     & 0.6$^{*}$ &  " &  " &  -  &  - &  \\ \noalign{\vskip 1.7mm} 
 
 & Gas disc        & 5.1 $\pm$ 0.3 & 10:00:18.255 & 2:12:42.635 & 0.3 $\pm$ 0.05 & 8 $\pm$ 4  & -  \\ 
ID13 & Stellar disc   & 3.34 $\pm$ 0.02 & 10:00:18.229$^{*}$ & 2:12:42.502$^{*}$ &  0.14$^{*}$ &  32$^{*}$ & \multirow{2}{*}{0.16} \\ 
& Stellar bulge     & 0.6$^{*}$ &  " &  " &  -  &  - &  \\ \noalign{\vskip 1.7mm} 
\hline
\end{tabular}
\end{table*}

\section{Results}\label{sec5:results}

\subsection{Surface Brightness best-fit models}\label{sec5:res_dysb}
\begin{figure*}
    \centering
    \includegraphics[width=0.99\linewidth]{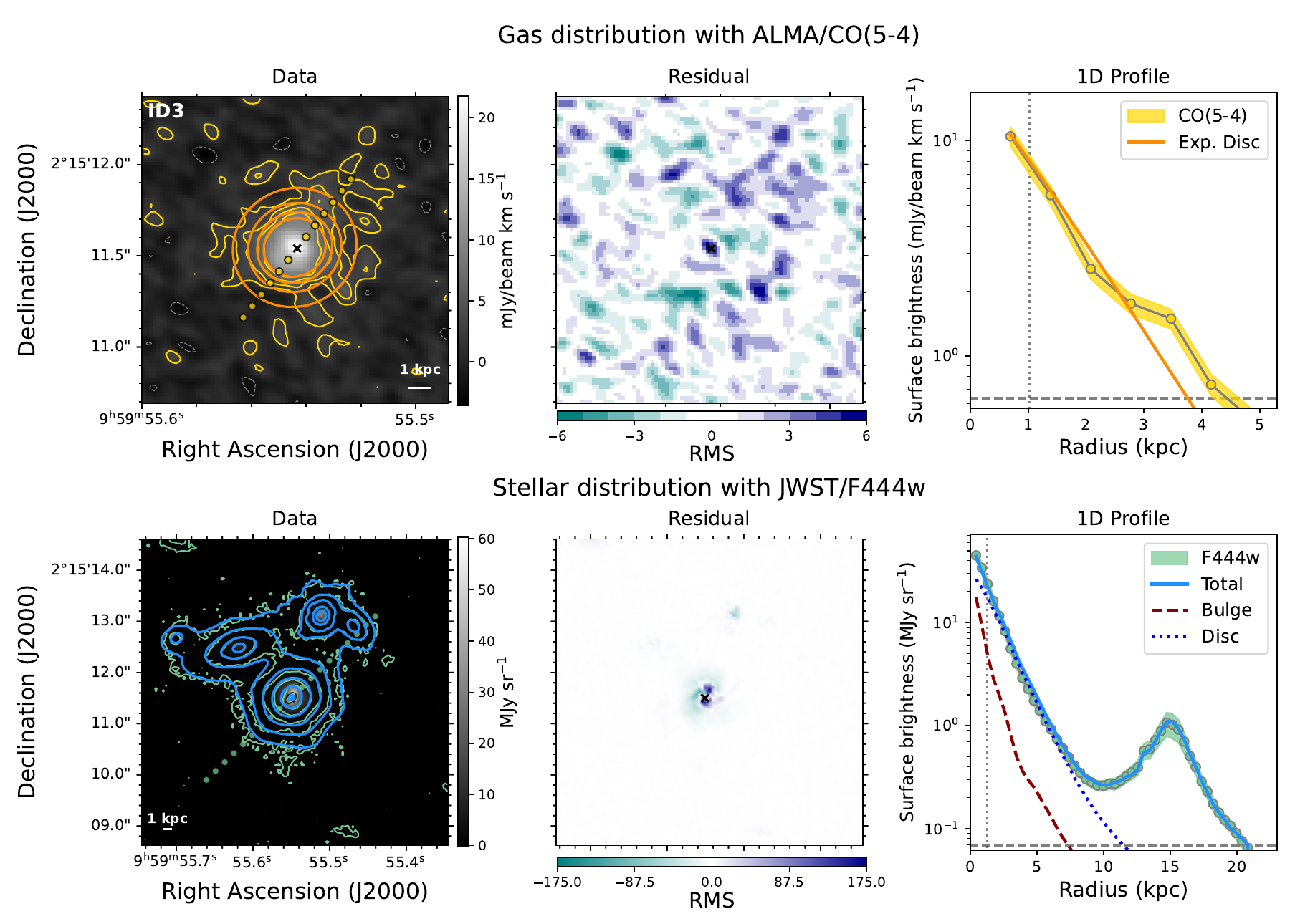}
    \caption[Surface brightness models of the gas and stellar distribution of ID3.]{Surface brightness models of the gas (top panels) and stellar distribution (bottom panels) of ID3. The panel description is the same as Figure~\ref{fig:sb_id1}. We note that there are foreground/background galaxies seen visible in the near-infrared (bottom left panel) that are fitted simultaneously to the target galaxy.}
    \label{fig:sb_id3}
\end{figure*}

\begin{figure*}
    \centering
    \includegraphics[width=0.99\linewidth]{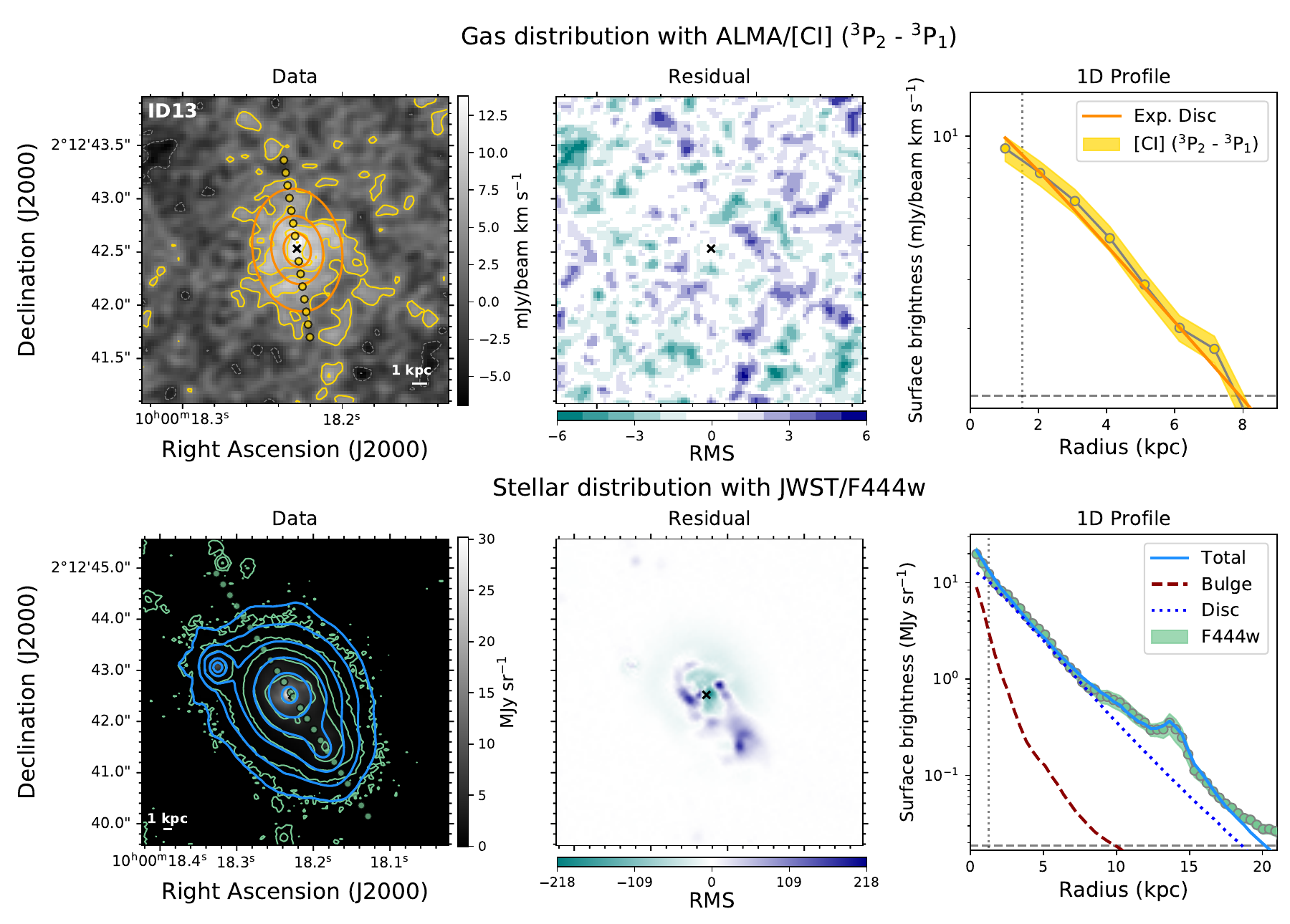}
    \caption[Surface brightness models of the gas and stellar distribution of ID13.]{Surface brightness models of the gas (top panels) and stellar distribution (bottom panels) of ID13. The panel description is the same as Figure~\ref{fig:sb_id1} and, similar to ID3, this galaxy is also surrounded by foreground/background galaxies that are fitted simultaneously.}
    \label{fig:sb_id13}
\end{figure*}

In Figures~\ref{fig:sb_id1}, ~\ref{fig:sb_id3} and ~\ref{fig:sb_id13}, we show the best-fit models for the gas and stellar surface brightness of ID1, ID3 and ID13, respectively. 
The corresponding corner plots are shown in Appendix~\ref{sec5:dysb_corner}. 
We report in Table~\ref{tab:dysb} the parameters defining the surface brightness for the gas disc, the stellar disc and the stellar bulge: the effective radius ($R_{\mathrm{eff}}$), the centre ($x_0$ and $y_0$), the ellipticity ($\epsilon$) and position angle (PA). 
Under the thin disc assumption, the ellipticity values of the stellar disc result in values of the inclination angles consistent with those used in \citet{rizzo23} to fit the gas kinematics. 
In Table~\ref{tab:dysb}, we also report the derived bulge-to-total luminosity ratio (B/T) for the stars.

In general, we obtain good fits, as can be confirmed by looking at both the 2D distribution and the 1D profiles.
We found that there are no obvious systematic residuals in the gas surface brightness fits; albeit, ID3 has a gas "tail" that is not well reproduced by the single exponential disc. However, this brightness is low and is located mostly on one side of the galaxy.
The stellar surface brightness profiles are very well reproduced by our multi-component models.

For the stellar component surface brightness fits, the maps of RMS-normalised residuals show much larger values than those in the gas component. 
There are a few reasons why this may be happening. 
i) The \textit{JWST} data have a higher angular resolution and a much higher signal-to-noise ratio than the ALMA data. 
This can boost the effect of any asymmetric feature. The opposite happens for the gas, as the lower resolution of the data has the effect of smoothing any inhomogeneities. 
ii) The stellar component is not fully axi-symmetric, as it contains features such as spiral arms and/or a bar, which are not accounted for in our model; this is best seen in ID1. 
iii) ID3 and ID13 have foreground/background galaxies that need to be fitted simultaneously and are out of the velocity range of the ALMA data; therefore, it introduces more uncertainty in the stellar surface brightness that is not present in the gas data.

\subsection{Mass models}\label{sec5:rotcur}

Before modelling the mass contributions to the rotation curve, we first obtain the circular speed from the observed rotation by applying the asymmetric drift correction described in Section~\ref{met:adc}.
The final rotation curves are shown in Figure~\ref{fig:fid} along with the best-fit fiducial models and in Figure~\ref{fig5:adc} with a comparison to the observed rotation velocities. We note that the correction is minimal, as the three galaxies analysed have low velocity dispersion profiles and high degrees of rotation support ($V/\sigma \gg 1$).

\subsubsection{Fiducial mass models}\label{sec5:fid}
 We show the results of the fiducial models for the three galaxies in Figure~\ref{fig:fid} and in Table~\ref{tab:fid}. The cornerplots for these fits are shown in Appendix~\ref{sec5:dydy}. As discussed in Section~\ref{sec5:modelling}, we tried to constrain the three main mass components: the total stellar mass, the total gas mass and the dark matter halo virial mass. Even though we have estimates of the stellar masses from SED fitting, the stellar masses might have systematic uncertainties of 0.3 dex \citep[e.g.][]{pacifici23, Choe_2025} and the possible differences in $\Upsilon$ due to different stellar populations in the bulge and disc are unknown. Moreover, the mass contribution of the bulge relative to the disc is very uncertain, considering that for the three targets, the bulges seem to be unresolved (see Section~\ref{sec5:res_dysb}). For the gas component, the emission line ratios and $\alpha_{\mathrm{CO}}$ are uncertain for the CO emission of ID1 and ID3. For ID13, we have a good estimate of the [CI] line ratio by assuming a typical gas excitation temperature, however, the typical values of $\alpha_{\mathrm{[CI]}}$ from the literature tend to be too high for the observed rotation curve, as discussed in Section~\ref{sec5:modelling}.

\begin{figure*}
    \centering
    \includegraphics[width=0.98\textwidth]{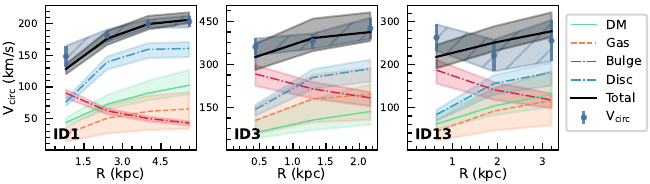}
    \caption[Fiducial mass models for selected ALPAKA galaxies.]{Fiducial mass models. The observed circular speed is shown in dark blue points and the best-fit model of the total contribution of all baryonic components and dark matter (DM) is shown as a solid black line. The individual components are shown separately: the stellar bulge and disc in red and blue dash-dotted lines, respectively; the gas disc in dashed orange lines; and the NFW dark matter halo in solid green lines. The uncertainties are shown as a shaded band around each component, data and model, indicating the corresponding 16th and 84th percentiles.}
    \label{fig:fid}
\end{figure*}

\begin{table*}
\caption{Best-fit parameters of our fiducial rotation curve decompositions. The free parameters of the fit are the total stellar mass ($M_{*}$), the gas normalisation ($\alpha_{\mathrm{CO/[CI]}} \times r_{l}$) and the baryonic fraction ($f_{\mathrm{bar}}$). The stellar bulge and stellar disc masses are calculated using the B/T ratio obtained with \texttt{GALFIT} assuming that the bulge has a mass-to-light ratio 1.4 times that of the disc. The total gas mass (M$_{\mathrm{gas}}$) is calculated by multiplying the CO/[CI] luminosity by the gas normalisation. The dark matter halo virial mass ($M_{200}$) is calculated from the f$_{\mathrm{bar}}$ fraction parameter. We note that the $c_{200}$ parameter is derived from the concentration-mass relation from \citep{dutton14} and fixed in our fit.}
\label{tab:fid}
\centering
\begin{tabular}{ccc ccc ccc}
    \hline\hline \noalign{\vskip 1mm}
     \multirow{3}{*}{ID} & 
\multirow{3}{*}{$\log \left( \frac{M_{\mathrm{*}}}{M_{\odot}} \right)$} & 
\multirow{3}{*}{$\log \left( \frac{M_{\mathrm{b}}}{M_{\odot}} \right)$} & 
\multirow{3}{*}{$\log \left( \frac{M_{\mathrm{d}}}{M_{\odot}} \right)$} & 
\multirow{3}{*}{$\log \left( \frac{M_{\mathrm{gas}}}{M_{\odot}} \right)$} & 
$\alpha_{\mathrm{CO/[CI]}} \times r_{l}$ & 
\multirow{3}{*}{$\log \left( \frac{M_{\mathrm{200}}}{M_{\odot}} \right)$} & 
\multirow{3}{*}{$\log {f_\mathrm{bar}}$} & 
\multirow{3}{*}{c$_{200}$} \\
    & & & & & \multirow{2}{*}{$ \left( \frac{M_\odot}{10^{10} \mathrm{K} \hspace{0.8mm} \mathrm{km} \hspace{0.8mm} \mathrm{s}^{-1}\mathrm{pc}^2} \right)$} & & & \\
    & & & & &  & & & \\   \hline \noalign{\vskip 1.5mm}
    
ID1 & 10.6$^{+0.1}_{-0.1}$ & 9.39 & 10.56 & 9.6$^{+0.3}_{-0.6}$ & 3.1$^{+3.0}_{-2.3}$ & 12.0$^{+0.6}_{-0.4}$ & $-1.3^{+0.4}_{-0.6}$ & 6.38  \\ \noalign{\vskip 1.7mm} 
ID3 & 10.9$^{+0.1}_{-0.1}$ & 10.34 & 10.75 & 10.2$^{+0.3}_{-0.5}$ & 1.0$^{+1.1}_{-0.7}$ & 13.5$^{+1.7}_{-1.3}$ & $-2.6^{+1.3}_{-1.7}$ & 3.71  \\ \noalign{\vskip 1.7mm} 
ID13 & 10.7$^{+0.1}_{-0.2}$ & 10.08 & 10.62 & 10.0$^{+0.4}_{-0.6}$ & 1.0$^{+1.6}_{-0.8}$ & 12.5$^{+1.0}_{-0.7}$ & $-1.7^{+0.7}_{-1.0}$ & 3.83  \\ \noalign{\vskip 1.7mm} 
\hline
\end{tabular}
\end{table*}

Similarly to the results in \citet{romanoliveira24}, we found that the total stellar masses are generally well constrained with small uncertainties, as seen in the corner plots (see Figures~\ref{fig:dydy_ID1_corner} -- \ref{fig:dydy_ID13_corner}). 
This is due to the fact that the stars dominate the gravitational potential in the inner region of the galaxies, and more specifically, the stellar bulge is the main contributor to this.
The total stellar masses derived from our fiducial fits are significantly different from those reported in Table~\ref{tab:data}, which were obtained from SED fitting, if the quoted errors are fair representations of the uncertainties. 
However, if we consider systematic errors of the order of 0.3 dex for the SED models \citep[see ][]{pacifici23, Choe_2025}, then the masses obtained for the three galaxies are borderline within the uncertainties. 
For ID1 in particular, the dynamical stellar mass is significantly higher than that expected from the SED fitting; however, the SED fitting used was obtained before the \textit{JWST} data became available. 
A preliminary SED fitting, including the \textit{JWST} bands, skews the SED mass to 2-3 times higher and closer to our dynamical estimate.

\begin{figure}
    \centering
    \includegraphics[width=0.9\columnwidth]{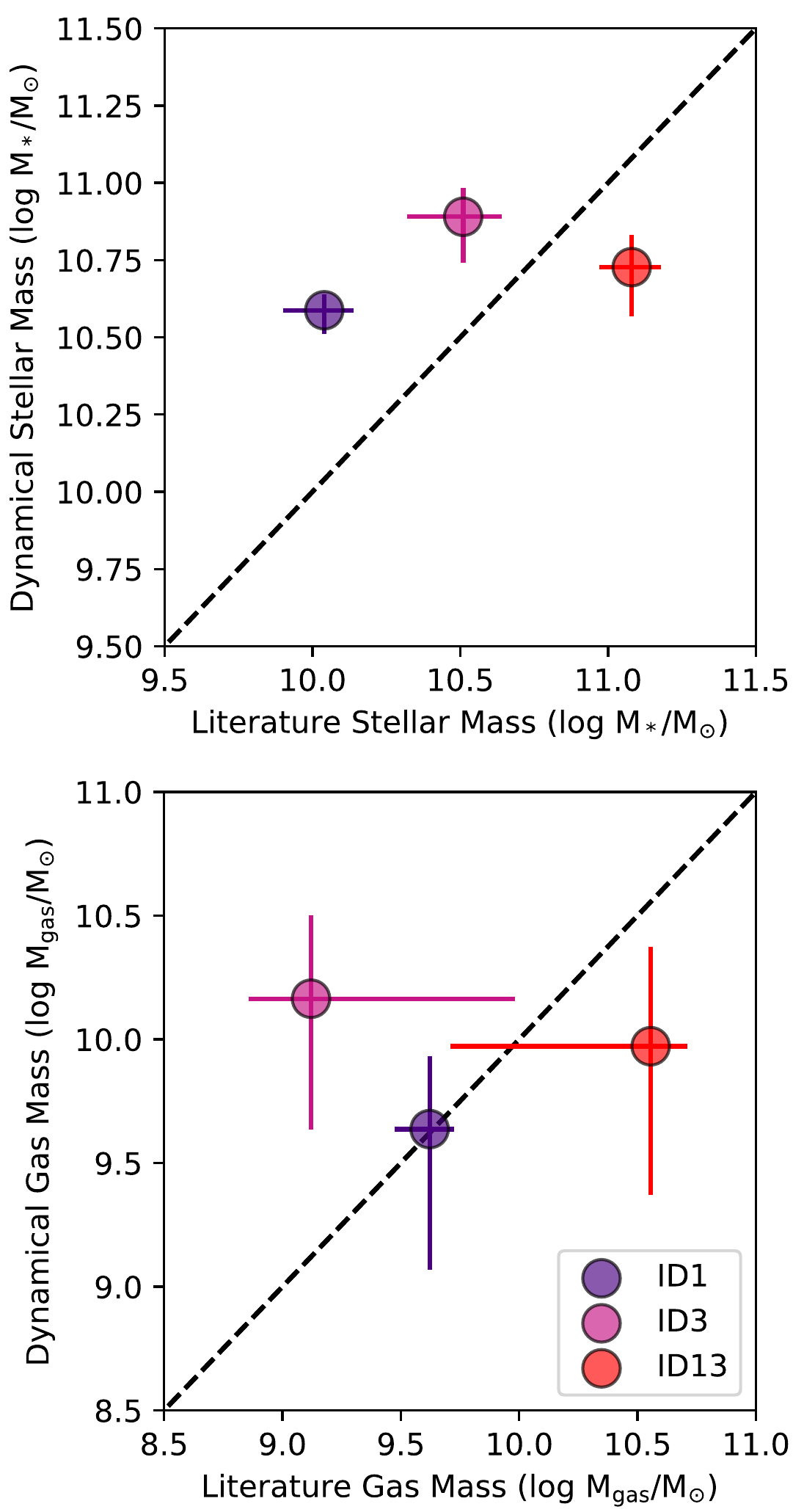}
    \caption[Comparison between dynamical and literature stellar and gas masses.]{Comparison of the expected values from the literature and the fiducial dynamical masses. Left panel: the dynamical stellar mass and errors come from our fiducial mass models, while the literature stellar mass and errors come from SED fitting and are reported in Table~\ref{tab:data}. Right panel: the dynamical gas mass and errors come from our fiducial mass models and we estimate the literature gas mass using Equation~\ref{eq:alpha}, we assume $\alpha_{\mathrm{CO/[CI]}}$ values of 3.6, 0.8 and 7 M$_\odot /10^{10} \mathrm{K} \hspace{0.8mm} \mathrm{km} \hspace{0.8mm} \mathrm{s}^{-1}\mathrm{pc}^2$ and line ratios of 0.83, 0.11 and 0.57 for ID1, ID3 and ID13, respectively. The errorbars are calculated with an estimated range as detailed in the text. In both panels we show the identity line as a dashed black line.}
    \label{fig:exp}
\end{figure}

Contrary to the stellar components, the parameters of the gas and dark matter components are not easily constrained, as one can infer from the posteriors (see Figures~\ref{fig:dydy_ID1_corner} -- \ref{fig:dydy_ID13_corner}).
This is because, in the outskirts of our galaxies, the gas disc and the dark matter halo contributions have very similar shapes. 
These contributions are then degenerate, as we are not probing the regions where dark matter dominates the gravitational potential.
By using the current data and assumptions, there is too little information to constrain the gas and dark matter masses and, therefore, to break this disc-halo degeneracy.

Even though we do not obtain precise constraints for the $\alpha_{\mathrm{CO/[CI]}} \times r_{l}$, the median values we found do not differ significantly from what was expected. If we divide the $\alpha_{\mathrm{CO/[CI]}} \times r_{l}$ shown in Table~\ref{tab:fid} by the expected line ratios from the literature (0.83, 0.11 and 0.57 for ID1, ID3 and ID13 respectively), the estimated $\alpha_{\mathrm{CO/[CI]}}$ values are 3.7, 9.1 and 1.8 M$_\odot /10^{10} \mathrm{K} \hspace{0.8mm} \mathrm{km} \hspace{0.8mm} \mathrm{s}^{-1}\mathrm{pc}^2$. For ID1 and ID13, these values are within the range of expected values in the literature; albeit, ID13 is on the lower end. 
For ID3, the $\alpha_{\mathrm{CO}}$ is quite high, but it might be more indicative that the line ratio of 0.11 for the CO~(5-4)/CO~(1-0) transitions is inappropriate, as the estimate for the line ratio of CO~(5-4) is based on very few objects (see Section~\ref{sec5:modelling}).
If we divide the measured $\alpha_{\mathrm{CO}} \times r_{l}$ of ID3 by the expected $\alpha_{\mathrm{CO}} = 0.8$ M$_\odot /10^{10} \mathrm{K} \hspace{0.8mm} \mathrm{km} \hspace{0.8mm} \mathrm{s}^{-1}\mathrm{pc}^2$ for starburst galaxies, we find that the luminosity line ratio would be 1.25.
Alternatively, if the $\alpha_{\mathrm{CO}}$ is closer to 4.0 M$_\odot /10^{10} \mathrm{K} \hspace{0.8mm} \mathrm{km} \hspace{0.8mm} \mathrm{s}^{-1}\mathrm{pc}^2$, then the line ratio would be 0.25.
The latter $\alpha_{\mathrm{CO}}$ might not be unreasonable for a starburst galaxy, as in \citet{dunne22} they show that an $\alpha_{\mathrm{CO}}$ of 4 M$_\odot /10^{10} \mathrm{K} \hspace{0.8mm} \mathrm{km} \hspace{0.8mm} \mathrm{s}^{-1}\mathrm{pc}^2$ is a near-universal average value for metal-rich galaxies. However, it is challenging to break the degeneracy between $\alpha_{\mathrm{CO}}$ and $r_l$ using the constraints in hand.

To summarise, in Figure~\ref{fig:exp} we show an overall comparison of the stellar and gas masses obtained from our fiducial mass models and the expected values from the literature. 
For the stars, we compare our fiducial stellar masses with the stellar masses and their uncertainties obtained from SED models given in Table~\ref{tab:data}. 
For the gas, we compare our fiducial gas mass with an estimate of the gas mass using Equation~\ref{eq:alpha} and the CO/[CI] luminosities reported in Table~\ref{tab:data}. 
These gas mass estimates are obtained using typical values in the literature for $\alpha_{\mathrm{CO/[CI]}}$ of 3.6, 0.8 and 7 M$_\odot /10^{10} \mathrm{K} \hspace{0.8mm} \mathrm{km} \hspace{0.8mm} \mathrm{s}^{-1}\mathrm{pc}^2$ and line ratios of 0.83, 0.11, and 0.57 for ID1, ID3, and ID13, respectively. 
To estimate the errorbars shown, we assume a range of possible values for the $\alpha_{\mathrm{CO/[CI]}}$: for ID1, as a main-sequence galaxy observed in a CO transition, we assume that the $\alpha_{\mathrm{CO}}$ can vary between 3 and 4 M$_\odot /10^{10} \mathrm{K} \hspace{0.8mm} \mathrm{km} \hspace{0.8mm} \mathrm{s}^{-1}\mathrm{pc}^2$; for ID3, a starburst galaxy observed in a CO transition, we assume that the $\alpha_{\mathrm{CO}}$ can vary between 0.8 and 4 M$_\odot /10^{10} \mathrm{K} \hspace{0.8mm} \mathrm{km} \hspace{0.8mm} \mathrm{s}^{-1}\mathrm{pc}^2$; for ID13 we assume that the $\alpha_{\rm [CI]}$ ranges between 1 and 10 M$_\odot /10^{10} \mathrm{K} \hspace{0.8mm} \mathrm{km} \hspace{0.8mm} \mathrm{s}^{-1}\mathrm{pc}^2$ (see discussion in Section~\ref{sec5:modelling}). For the errors regarding the line ratios, we use the errors reported in \citet{boogaard20} for ID1 and ID3 and we assume no errors in the line ratio of ID13.

Regarding dark matter, as mentioned in Section \ref{sec5:modelling}, the virial mass ($M_{200}$) estimate is derived from the baryonic fraction parameters, whose posteriors can be seen in Figures~\ref{fig:dydy_ID1_corner} -- \ref{fig:dydy_ID13_corner}. 
These posterior distributions do not show clear peaks, but they seem to converge close to the edge of the prior, which corresponds to the cosmological baryon fraction, suggesting that these galaxies are baryon dominated. 
This is partially due to the fact that we are not probing the regions where dark matter dominates the gravitational potential.

Examining the fiducial mass models, we find that they reproduce the overall shape of the rotation curves for all three galaxies. 
However, in each case, the models systematically underestimate the circular velocity in the central regions. 
This effect is most evident at the innermost radii of ID1 and ID13, where the predicted velocities tend to fall toward the lower edge of the observational uncertainties.
Aside from the possibility that we are overestimating the circular speed in the inner regions of the galaxies (see Section \ref{sec5:discussion}), there could be physical reasons that we are not accounting for in our fiducial models.
Therefore, in Section~\ref{sec5:under}, we consider alternative scenarios that could explain this discrepancy, including the contributions of supermassive black holes, larger differences in the mass-to-light ratios of the bulge and the disc, and the effects of dust obscuration.

\subsubsection{Models with the baryonic parameters fixed}\label{sec5:dm}

Our fiducial mass models for the three ALPAKA galaxies led to the conclusion that we do not obtain a strong constraint on the dark matter halo virial mass and it would not be possible to constrain the NFW concentration parameter even if we let it free, due to the degeneracy with the gas component and the observational limit of this dataset.
We remind the reader that, as shown in Table~\ref{tab:data}, these three galaxies in the ALPAKA survey are observed in CO/[CI] with around two fully independent resolution elements per galaxy side, while we are trying to constrain three independent parameters with the fiducial model: the total stellar mass, the total gas mass and the dark matter mass.
A different strategy is to fix the stellar and gas masses of the three targets to the values obtained using other methods, which is what we adopt in this section.

For the gas masses, as mentioned in Section~\ref{sec5:modelling}, we assume that the CO luminosity-to-H2 mass conversion factor ($\alpha_{\mathrm{CO}}$) is 3.6 and 0.8 M$_\odot /10^{10} \mathrm{K} \hspace{0.8mm} \mathrm{km} \hspace{0.8mm} \mathrm{s}^{-1}\mathrm{pc}^2$ for the main-sequence galaxy ID1 and for the starburst galaxy ID3, respectively \citep{daddi10a, kirkpatrick19}. 
We also assume that the CO luminosity line ratios from \citet{boogaard20} of 0.83 for the luminosity line ratio of CO~(2-1)/CO~(1-0) for ID1 and 0.11 for the line ratio of CO~(5-4)/CO~(1-0) for ID3.

On the other hand, the gas emission of ID13 is traced with the [CI]~($^3$P$_2$ - $^3$P$_1$) emission, which presents some caveats that are explained in Section~\ref{sec5:modelling}. The typical values found for the [CI] luminosity to gas mass conversion factor in the literature tend to be around 7 M$_\odot /10^{10} \mathrm{K} \hspace{0.8mm} \mathrm{km} \hspace{0.8mm} \mathrm{s}^{-1}\mathrm{pc}^2$; however, they can range from 1 to 10 M$_\odot /10^{10} \mathrm{K} \hspace{0.8mm} \mathrm{km} \hspace{0.8mm} \mathrm{s}^{-1}\mathrm{pc}^2$.
If we assume a gas excitation temperature of $30\,{\rm K}$ as suggested by previous works \citep[e.g.][]{walter11, valentino20b, henriquezbrocal22}, we obtain a line ratio (L'$_{\mathrm{[CI] ^3 P_2 - ^3 P _1}}$/L'$_{\mathrm{[CI] ^3 P_1 - ^3 P _0}}$) of 0.57. This gives us a total molecular gas of log M$_{\mathrm{gas}}$/M$_{\odot}$ = 10.6 when assuming the typical value of $\alpha_{\mathrm{[CI]}}$ = 7 M$_\odot /10^{10} \mathrm{K} \hspace{0.8mm} \mathrm{km} \hspace{0.8mm} \mathrm{s}^{-1}\mathrm{pc}^2$, or a total molecular gas of log M$_{\mathrm{gas}}$/M$_{\odot}$ = 9.7 when assuming a conservative $\alpha_{\mathrm{[CI]}}$ = 1 at the lower end of the expected $\alpha_{\mathrm{[CI]}}$ values. 
Translating this to the velocity contribution to the rotation curve, it would mean that the gas disc would contribute at $3.2 \kpc$ (the last point of the rotation curve) with $227 \kms$ or $86 \kms$ for $\alpha_{\mathrm{[CI]}}$ of 7 and 1 M$_\odot /10^{10} \mathrm{K} \hspace{0.8mm} \mathrm{km} \hspace{0.8mm} \mathrm{s}^{-1}\mathrm{pc}^2$, respectively. 
The former value is physically impossible, as the observed rotation at $3.2\kpc$ is $\approx 255\kms$, and it needs to accommodate both the gas disc and the stellar disc, contributing at least $200\kms$ when considering the stellar mass of ID13 reported in Table~\ref{tab:data} and the bulge-to-total luminosity ratio in Table~\ref{tab:dysb}. Therefore, we kept the conservative $\alpha_{\mathrm{[CI]}} = 1$ M$_\odot /10^{10} \mathrm{K} \hspace{0.8mm} \mathrm{km} \hspace{0.8mm} \mathrm{s}^{-1}\mathrm{pc}^2$ to fix the gas mass of ID13.

For the stellar masses, we use the values obtained from SED models for ID1 and ID3 reported in \citet{rizzo23} and shown in Table~\ref{tab:data}. 
For ID13, we assume a more conservative stellar mass: we take the SED fit stellar mass from Table~\ref{tab:data} and subtract 0.3 dex, which is the suggested value of systematic uncertainties in SED models \citep{pacifici23}. 
We made this choice because, if we use the stellar mass in Table~\ref{tab:data}, the contributions from the stellar bulge and disc alone already overpredict the rotation in the galaxy.
In all cases, we kept the assumption of a difference in $\Upsilon$ of 1.4 between the bulge and disc components, and we maintained the shape of the surface density of the gas disc, stellar disc, and stellar bulge as obtained from the surface brightness fitting and already used in the fiducial models.
We thus obtain mass models in which the baryonic components are completely fixed and the parameters for the NFW dark matter halo (i.e., $M_{200}$ and concentration) are allowed to vary.

\begin{figure*}
    \centering
    \includegraphics[width=\textwidth]{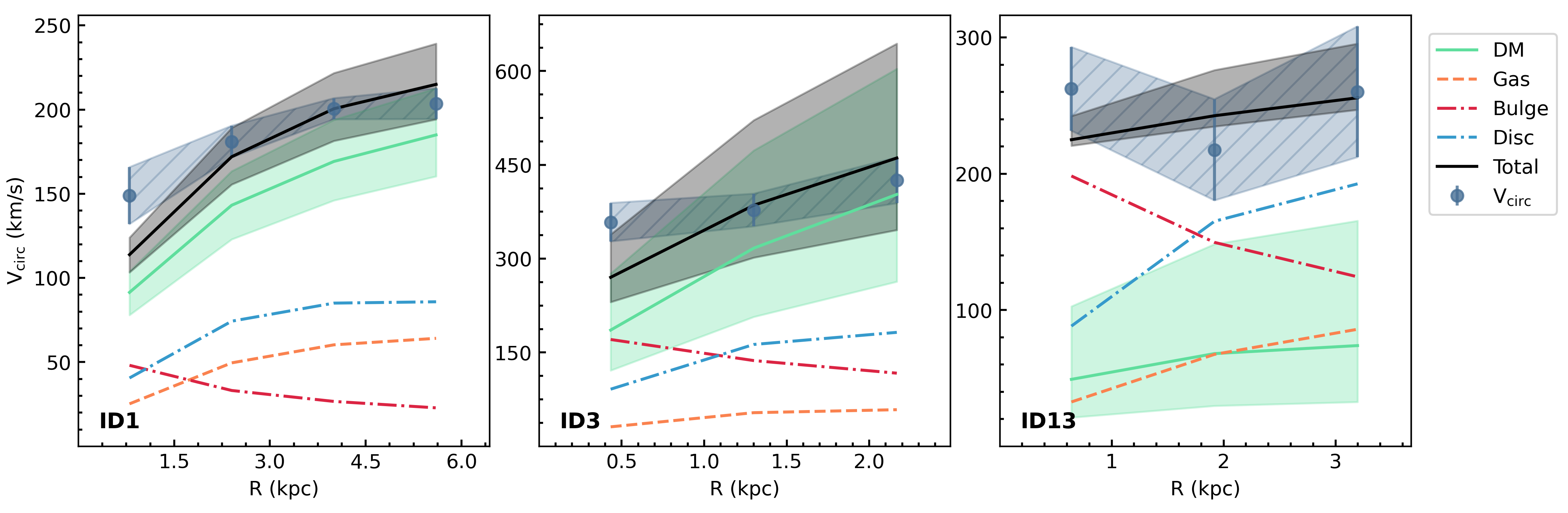}
    \caption[Mass models with the baryonic parameters fixed.]{Mass models with fixed baryonic parameters and free dark matter parameters (NFW mass and concentration). We use values from the literature for the $\alpha_{\mathrm{CO}/\mathrm{[CI]}}$, $r_l$ and for the total stellar mass, which allow us to fix the normalisations of the circular speed contributions of gas and stars. The shapes of the circular speed contributions are kept to the same as those derived from the surface brightness and used in the fiducial models shown in Figure~\ref{fig:fid}.}
    \label{fig:alt}
\end{figure*}

\begin{table}
\caption{Mass models with the baryonic parameters fixed. In this fit, we kept all the baryonic components (stellar disc, stellar bulge and gas) fixed with prior estimates in the literature and let the dark matter NFW halo mass and concentration free to vary. We note that the dark matter halo mass obtained for ID13 (signalled with an asterisk) is lower than that of the baryons, therefore physically unfeasible.}
\label{tab:dmmod}
\centering
\begin{tabular}{ccc}
    \hline\hline \noalign{\vskip 1.2mm}
    ID & $\log \left( \frac{M_{200}}{M_\odot} \right ) $ & $c_{200}$ \\   \hline \noalign{\vskip 1.8mm}
1 & 12.0$^{+0.2}_{-0.1}$ & 18.0$^{+1.4}_{-2.9}$ \\ \noalign{\vskip 1.7mm} 
3 & 14.0$^{+0.5}_{-0.4}$ & 14.8$^{+3.7}_{-4.1}$ \\ \noalign{\vskip 1.7mm} 
13 & *10.4$^{+1.1}_{-0.9}$ & 9.4$^{+7.4}_{-5.2}$ \\ \noalign{\vskip 1.7mm} 
\hline
\end{tabular}
\end{table}

In these models, we use a flat log prior for the NFW halo mass (M$_{200}$) that range, from 10.5 to 15 $\log(M_{200}/M_\odot)$ and a flat prior limited from 2.5 to 20 for the NFW halo concentration ($c_{200}$). The concentration prior range roughly represents the possible concentrations expected for NFW dark matter halos in the local Universe and up to $z=4$ with virial masses between 10$^{11}$ and 10$^{15}$ M$_\odot$ \citep{dutton14, ludlow14}.

We show these new models in Figure~\ref{fig:alt} and present the results in Table~\ref{tab:dmmod}, where it is immediately apparent that, for ID1 and ID3, the models struggle to reproduce the observed rotation curve in a manner similar to that of the fiducial models.
The main reason for the results for ID1 and ID3 is that the almost flat shape of the observed rotation curves cannot be reproduced by an NFW halo with a concentration value expected in CDM cosmology. 
We experimented by leaving the concentration free with a much larger prior range and we found that the best model tends towards concentration values upwards of 50, above a 5-$\sigma$ level of statistical significance from the expected concentration-mass relation in \citet{dutton14}.
Unlike ID1 and ID3, which tend to be dark matter dominated in this fit, ID13 is dominated by the stellar component.
We note that this fit is very similar to the fiducial one, given that we are assuming a stellar mass similar to that found with the fiducial models. 
This way, the only change occurs by keeping the gas mass fixed and allowing the dark matter halo mass to vary freely. 
This, however, leads to a small dark matter halo mass, as we do not impose the cosmological baryonic fraction, while the dark matter halo concentration has large uncertainties of $\gtrsim 50\%$. 
We note, again, that this fit only works when assuming conservative contributions of both the stellar and gas masses.

In conclusion, with the current data and assuming that we know the parameters describing the observed baryonic components, we are capable of obtaining estimates of the dark matter halo virial masses with significantly smaller errors than in our fiducial models (Table~\ref{tab:dmmod}).
However, for both ID3 and ID13 the obtained values do not appear realistic. 
Moreover, it is still not possible to determine the concentrations of the dark matter halos and we are not capable of describing the observed rotation curve in its innermost point at a 1$\sigma$ level. 
If anything, the data suggest the need for  high values of the concentration parameter. 
The higher concentration would allow for the NFW halo to have more mass in the centre and, therefore, a higher dark-matter contribution to the inner velocity point in the rotation curve of the galaxies. 
Alternatively, we can mitigate the need for a highly concentrated dark matter component with different assumptions for the baryonic component, as discussed in the next Section.

\section{Discussion}\label{sec5:discussion}

\subsection{Missing mass in the center}

In Section \ref{sec5:res_dysb}, we show that reproducing the rotation curves of our three targets requires fine-tuning of the mass model and allowing for a significant contribution from the bulge in the inner regions of the galaxies. 
We proposed fiducial models that account for the strong matter contributions in the central regions by maximising the bulge luminosities within the uncertainties of the surface brightness and including differences in the mass-to-light ratios of the bulge and the disc. 
In this Section, we explore other possible options for the mass models that could help explain the discrepancy between the observed inner rotation and the expected values.

\subsubsection{Sensitivity of the kinematic models to bulge contribution}

\begin{figure*}
    \centering
    \includegraphics[width=0.98\textwidth]{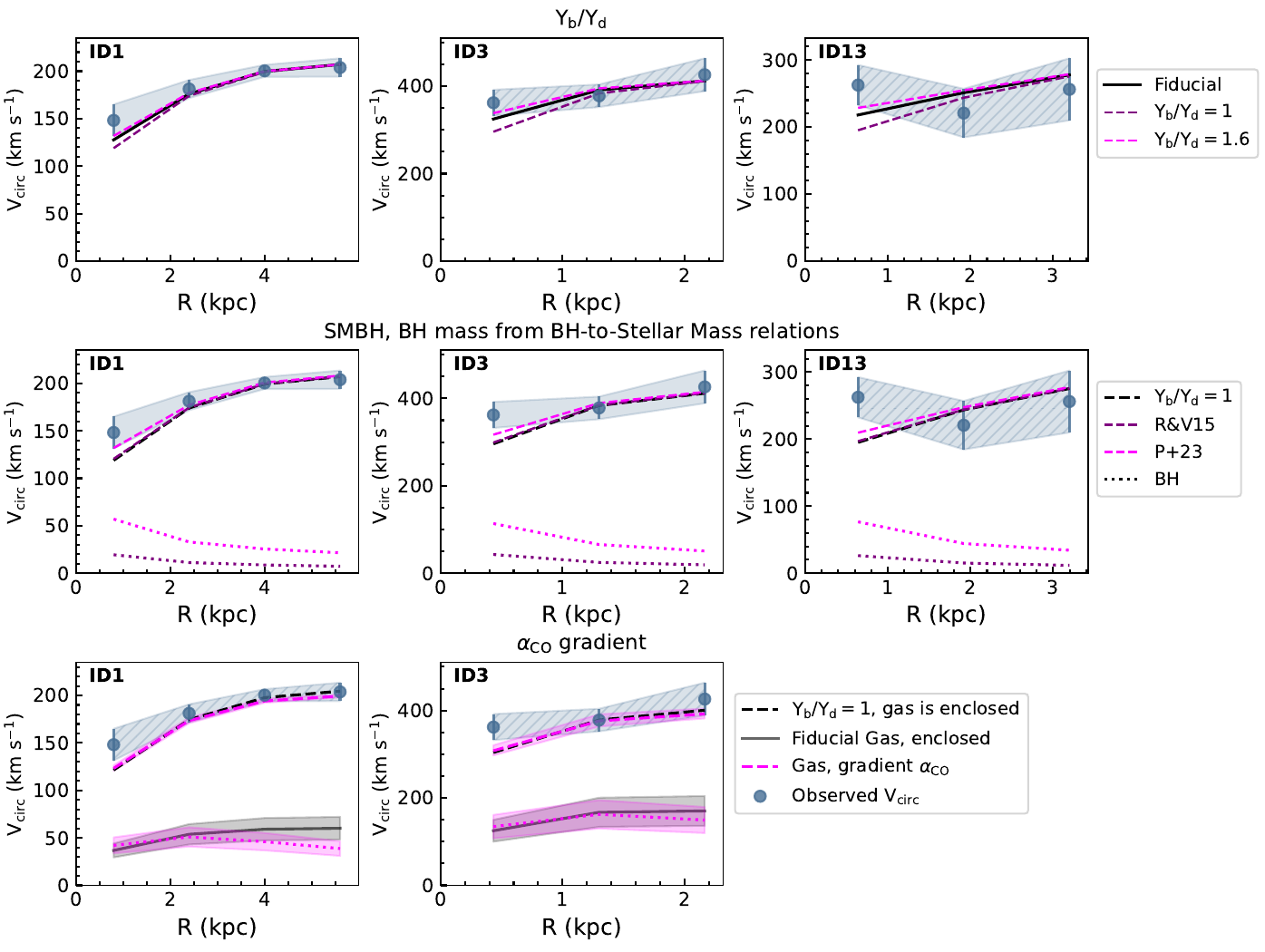}
    \caption[Effect of different physical assumptions on the mass models.]{Effect of different physical assumptions on the mass models. Top panels: Effect of differences in the mass-to-light ratios of the bulge and the disc in the rotation curves of the ALPAKA galaxies (the columns show the results for ID1, ID3 and ID13 from left to right). The fiducial model assumes that the $\Upsilon$ of the bulge is 1.4 times that of the disc. We show the tests with a $\Upsilon$ difference of 1.6 (in magenta) and no difference (in purple). Middle panels: Effect of the presence of a supermassive black hole. We show the effect of a supermassive black holes that follows the black hole-stellar mass relation for local galaxies \citep[in purple, ][R\&V15]{reines15} and for high-$z$ galaxies \citep[in magenta, ][P+23]{pacucci23}. Bottom panels: The approximate effect of a positive metallicity gradient on the $\alpha_{\mathrm{CO}}$ and, therefore, on the fiducial model. We do not show ID13 as it is observed in [CI]. 
    }
    \label{fig:alt_all}
\end{figure*}

When we assume that the bulge and stellar disc have the same mass-to-light ratio ($\Upsilon_{\mathrm{b}}/\Upsilon_{\mathrm{d}} = 1$; dashed purple line in Fig.~\ref{fig:alt_all}), the resulting model underestimates the observed rotation velocity at the innermost data point. 
To investigate whether this discrepancy can be mitigated, we constructed kinematic models, i.e.\ mock emission line datacubes.
Essentially we took the the kinematic parameters defining the best-fit models in \citet{rizzo23} and changed the first points of the rotation curves, assigning the values we obtained from the above dynamical models with $\Upsilon_{\mathrm{b}}/\Upsilon_{\mathrm{d}} = 1$.
These values are $120 \kms$, $277\kms$, and $189 \kms$ for IDs 1, 3, and 13, respectively.
We then computed the $\chi^2$ values for each of these models and compared them to the best-fit $\chi^2$ values from \citet{rizzo23} using the Akaike Information Criterion (AIC). This comparison indicates that, for ID1, the two models are not significantly different, suggesting that the first data point may be biased toward higher velocities. 
In contrast, for IDs 3 and 13, the AIC differences are substantially larger ($\gtrsim$ 1000), implying that the fixed models provide a significantly poorer fit compared to the best-fit models reported in \citet{rizzo23}.

\subsubsection{Massive bulges}\label{sec5:under}

In this Section, we explore potential reasons for the need for a different mass-to-light ratio for the bulge and the disc. At $z = 0$, assuming different mass-to-light ratios for the bulge and the disc is standard \citep{lelli16b}. This assumption is based on the expected difference in the age/metallicity of the older stellar populations in the bulge, which can drive a difference in colours and, therefore, in the mass-to-light ratios.
If the bulges in the ALPAKA galaxies are sufficiently old compared to the disc, then the $\Upsilon$ difference could be similar to those of local galaxies where the $\Upsilon_{\mathrm{b}}/\Upsilon_{\mathrm{d}}=1.4$ is a common assumption based on stellar population synthesis models \citep{schombert14, lelli16b}.

The main limitation involved in the analysis of the difference in $\Upsilon$ in the ALPAKA galaxies is the challenge of obtaining constraints on the star formation history. 
However, considering the age of the Universe at the three redshifts of these galaxies, we can estimate the maximum difference between the mass-to-light ratios of the bulge and disc.
\citet{schombert19} shows that the difference in $\Upsilon$ between a 3 Gyr old stellar population with solar metallicity and a younger population of $\sim$ 300 Myr with metallicity of [Fe/H] = $-0.5$ can reach 1.6.
Therefore, based on these values, we test a scenario in which the bulge and the disc have a $\Upsilon$ difference of 1.6, to represent the extreme scenario of an old bulge and a constant star forming disc. This is not a fit, but only a rescaling of the bulge and disc contributions.
We show this in the top row of Figure~\ref{fig:alt_all}, where we compare the fiducial model with $\Upsilon_{\mathrm{b}}/\Upsilon_{\mathrm{d}}=1.6$. 
We see that in all cases, the higher difference improves the fit in the inner region, as it tends to boost the central contribution of the stars. Whether values of $\Upsilon_{\mathrm{b}}/\Upsilon_{\mathrm{d}}$ above one are realistic for the stellar populations of these galaxies remains, however, an open question.

Another way of viewing the needed difference in the mass-to-light ratios of the bulge and the disc is that, instead of being driven by the stellar populations, it is driven by dust obscuration.
Despite these galaxies being observed in the rest-frame near infrared, the central light of the galaxies may still be significantly attenuated \citep[e.g.,][]{polletta24}, which leads to two distinct effects: i) determining the structure of the bulge (namely the effective radius) may be difficult, and ii) the total intensity of the bulge may be underestimated. Both of these could explain why the stellar surface brightness fits tended to underestimate the bulge contribution (see Section~\ref{sec5:res_dysb}).
To quantify the effect of dust attenuation, we assumed a total-to-selective extinction ratio $R_{\mathrm{V}}$ of 4.05 and a reddening curve $k(\lambda) = 2.659(-1.857 + 1.040/\lambda) + R_V$ \citep{calzetti00}.
The relation between observed and intrinsic luminosity is given by
\begin{equation}
    L_{\mathrm{obs}}(\lambda) = L_{\mathrm{int}}(\lambda) \times 10^{-0.4 A_{\lambda}},
\end{equation}
where $A_{\lambda}$ is related to the reddening law by $A_{\lambda} = k(\lambda) A_{V}/R_{V}$.
To derive the reddening difference between bulge and disc ($ \Delta A_{\mathrm{V}}$), we can rewrite this as
\begin{equation}
        \Delta A_{\mathrm{V}} = A_{\mathrm{V, d}} - A_{\mathrm{V, b}} = - 2.5 \frac{R_V}{k(\lambda)} \log \left( \frac{L_{\mathrm{d, int}}}{L_{\mathrm{d, obs}}} \frac{L_{\mathrm{b, obs}}}{L_{\mathrm{b, int}}}  \right), 
\end{equation}
where $L_{\mathrm{j, obs}}$ and $L_{\mathrm{j, int}}$ are, respectively, the observed and intrinsic luminosities of component $j$ (bulge or disc).

We note that the bulge/disc ratio of observed luminosities ($L_{\mathrm{b, obs}}/L_{\mathrm{d, obs}}$) is directly obtained from the surface brightness fits and the ratio of intrinsic luminosities can be inferred from the dynamical stellar mass ratios obtained with the fiducial fit and by assuming $\Upsilon_{\rm b}$ and $\Upsilon_{\rm d}$. Here, we test the assumption that the $\Upsilon$ is the same for the bulge and the disc and that the reddening is driving the difference between the observed luminosity ratio and the mass ratio necessary to reproduce the rotation curve.

Following this, we found that the three ALPAKA bulges must have, on average, extinctions of A$_V$ of $\approx$ 1.7, 3.0 and 1.7 mag above that of the disc, respectively, for ID1, ID3, and ID13. 
We note that these results can vary under different assumptions of the dust model, as they can depend on the spectral types and physical properties of the galaxies \citep{kriek13}. 
However, if we assume that the dust is mostly concentrated close to the bulge, then an inner A$_V$ of $\sim$ 2 mag is in line with typical dusty galaxies at z $\approx 2$ \citep[e.g.][]{dacunha15}.
Moreover, some dusty star forming galaxies at $z = 1.7 - 3$ observed with \textit{JWST} have been found to have strong colour gradients and heavily obscured bulges \citep{miller22, hodge24, polletta24}.  
We conclude that considering a larger value for the mass-to-light ratio of the bulge with respect to that of the disc can be a realistic choice.

\subsubsection{Alternative to massive bulges as main contributors to the inner rotation velocities}\label{sec5:under}

Another way to boost the rotation velocity in the innermost region is to consider the presence of a supermassive black hole.
Despite the fact that we use data that have not been taken to resolve the sphere of influence, the addition of a SMBH can potentially contribute to the inner rotation curve \citep[e.g.,][]{ruffa23, zhang24}.
In the second row of Figure~\ref{fig:alt_all}, we show the effect of a SMBH on the mass models by comparing a black hole that follows the black hole--stellar-mass relation for local galaxies from \citet{reines15} and the relation for galaxies at $z = 4 - 7$ from \citet{pacucci23}.
To explore this scenario, we compare the mass models including SMBHs with models similar to our fiducial models, but without the assumption of the $\Upsilon$ difference between bulge and disc; we call this the baseline model. 
We note that we are not refitting the rotation curves; rather, we are only adding a black hole component to the best-fit rotation curve.

We found that, in all three cases, a supermassive black hole that follows the local relation from \citet{reines15} does not produce results that are significantly different from the baseline model.
On the other hand, when using the relation from \citet{pacucci23}, which implies a black hole mass of approximately 1.5\% of the total stellar mass instead of only $\sim$0.25\% as in \citet{reines15}, we find a significant increase in the central velocity of the model and an overall change in the shape of the rotation curve. This is enough to bring the central velocity within the observational uncertainties for ID1; however, it is still not sufficient to reproduce the inner rotation in the other two galaxies. If we were to bring the mass models to overlap with the central observed velocity at a $1 \sigma$ level, then the black hole velocity contributions to the inner points should be $\approx 55$, 150, $110\kms$, which translate to SMBH masses of 1.5\%, 3.0\%, and 4.4\% of the total stellar mass for ID1, ID3 and ID13, respectively.

The high black hole to stellar mass ratios of high-$z$ galaxies are an ongoing debate in the literature, as many new \textit{JWST} observations suggest the presence of overly massive black holes when compared to the local relation \citep[e.g.,][]{goulding23, furtak24}. 
Although the presence of a black hole does not seem to be enough to reproduce the rotation curves of our targets, our results are promising from another perspective, as they show that with a resolution 3 to 4 times better than the current data, we could attempt to constrain SMBH masses at $z>0$.

Lastly, another possibility is the existence of a radial gradient in the CO luminosity to gas mass conversion factor, which has been suggested in previous studies of nearby galaxies \citep[e.g.,][]{sandstrom13, bigiel20}.
This is not straightforward to implement on the mass decomposition models, as the assumption that the gas is distributed in an exponential disc is no longer valid if we apply a radial variation of the $\alpha_{\mathrm{CO}}$.

However, we can conduct a simple experiment in which we derive the mass density of the galaxies with a radially varying $\alpha_{\mathrm{CO}}$ and estimate the circular velocity at each radius from the enclosed mass, V$_{\mathrm{c}}^2 = M_{\mathrm{enc}} G / R $. We note that this implicitly assumes that the mass is distributed in a spherical geometry; however, the circular velocity of a mass component distributed in a sphere is at most 20\% different from its circular velocity as if it were distributed in a disc \citep{binney08}.

In order for the gas mass distribution to be higher in the centre than in the outskirts, the $\alpha_{\mathrm{CO}}$ factor gradient needs to be negative. The $\alpha_{\mathrm{CO}}$ factor depends on the physical conditions of the gas (e.g., density, temperature and gas-phase metallicity) and it usually has a positive radial gradient in the galaxies where it has been observed \citep[e.g.,][]{sandstrom13, bigiel20}. This is because the $\alpha_{\mathrm{CO}}$ is lower in high-metallicity regions, which are usually found in the centre of galaxies.
For this gradient to explain the inner velocity discrepancy, the galaxies should have a large positive metallicity gradient. This is uncommon in the local Universe; however, at high-$z$, such gradients are possible. For instance, some works found positive and flat metallicity gradients at z $>$ 1 \citep{jones13, troncoso14, leethochawalit16, wang22, venturi24}. To maximise the gas contribution in the central regions, we assume the metallicity gradient of the galaxy YD4, part of the proto-cluster 2744-z7p9OD at z $\sim$ 7.9, which was reported in \citet{venturi24} to have a metallicity gradient ($\nabla_{\mathrm{r}} \log{\mathrm{Z}}$) of $+0.14 \pm 0.16$ dex kpc$^{-1}$.
This is the best example of a positive metallicity gradient in a high-$z$ galaxy, albeit \citet{troncoso14} also reports similar gradients in galaxies at z=$3-4$.
We then consider an upper-limit metallicity gradient of +0.3~dex~kpc$^{-1}$, corresponding to 1-\(\sigma\) above the median value measured for YD4, and compute for each galaxy the associated change in $\alpha_{\mathrm{CO}}$. Using the relation between $\alpha_{\mathrm{CO}}$ and metallicity from \citet{sandstrom13}, we convert this metallicity gradient into a radial $\alpha_{\mathrm{CO}}$ gradient. This yields $\nabla_{\mathrm{r}} \log \alpha_{\mathrm{CO}} = -0.072~\mathrm{dex~kpc}^{-1}$, which we adopt as the slope of the $\alpha_{\mathrm{CO}}$ radial profile. For the intercept, we assumed that the $\alpha_{\mathrm{CO}}$ of the inner point of the rotation curve to be the same as the baseline, so we can compare the change of the gas circular velocity with the baseline model.

We show this comparison in the bottom row of Figure~\ref{fig:alt_all}, where we present the total baseline model, the gas velocity from the baseline model and the gas velocity considering such an extreme $\alpha_{\mathrm{CO}}$ gradient.
In this case, to keep the comparison fair, we use the same approach of enclosed mass for the gas in the baseline model (that is, an exponential sphere).
We found that the effect of this $\alpha_{\mathrm{CO}}$ gradient only slightly contributes to the increase of the circular velocities in the innermost regions. Therefore, we conclude that such gradients are unlikely to alleviate the tension between the shapes of the models and the observed rotation curves. 
Moreover, this experiment tends to go in a somewhat opposite direction to the previous one: the $\Upsilon$ difference requires higher stellar metallicity in the central parts to have an old red bulge, while the $\alpha_{\mathrm{CO}}$ gradient solution requires a lower gas phase metallicity towards the centre.
This may indicate that only one of the solutions is viable or that stars and gas have significantly different metallicity gradients.
We also note that we forego this analysis for ID13, as the correlation between the $\alpha_{\mathrm{[CI]}}$ and the metallicity gradient is not well established.

Fine tuning the assumptions about the baryonic components was necessary to reproduce the data in its details. We found that plausible physical reasons that can play a role in maximising the bulge contributions are that these bulges have an older population than the disc or that their luminosities are more attenuated by dust than the disc is. The presence of supermassive black holes with masses on the order of 1\% of the stellar mass can also contribute to alleviating the discrepancies. 
The $\alpha_{\mathrm{CO}}$ gradient is an unlikely choice, as it depends on large positive metallicity gradients that are uncommon, even with a large positive metallicity gradient, it does not seem to affect the rotation curves significantly.
In the end, the most likely scenario is that some of these effects are concurrently influencing the observed velocities in different proportions. 
On a positive note, even in the most extreme assumptions laid out in this Section, we only see a change of a few tens $\kms$ in the total circular speed of the fiducial and baseline models, implying that the general results we found from our fiducial models are robust.

\subsubsection{Alternative to the assumed dark matter distributions}
\label{sec:alternatives}

The dynamical models presented so far have been constructed under the assumption that dark matter follows an NFW density profile. However, alternative halo models have been proposed, and although we do not explore them quantitatively here, they are worth briefly discussing.
One possibility is that the dark matter halo profile has been modified by adiabatic contraction due to the condensation of baryons in the central regions of galaxies \citep[e.g.,][]{Blumenthal_1986, gnedin_2004, Hussein_2025}. As gas cools and settles into the galactic disc and bulge, it deepens the gravitational potential well, causing the dark matter to contract and become more centrally concentrated. This process can steepen the inner slope of the halo relative to the NFW profile \citep{li2022}.

Another possibility arises from dark matter models that are alternatives to CDM. 
Particle physicists have proposed a variety of dark matter candidates in the dark sector that can lead to departures from the NFW shape \citep{Tulin_2018}. In particular, self-interacting dark matter (SIDM) models predict a wide diversity of halo structures, ranging from cored to even more cuspy profiles than NFW \citep[e.g.,][]{despali19, correa25}. Given the limited number statistics of our sample, we defer a detailed investigation of these alternative models to future work, when improved constraints on the baryonic distributions will also be available (e.g., from the ALMA–CONDOR sample\footnote{https://almascience.eso.org/alma-data/lp}).

\subsection{Dark matter fractions}

In recent years, several studies based on warm gas observations have estimated the dark matter fraction within the effective radius, $f_{\mathrm{DM}}(< R_{\mathrm{eff}})$, comparing it with simulations and with other galaxy properties, such as stellar surface density and star formation rate \citep{genzel17, bouche22, puglisi23, degraaff24}. 
The relatively low dark matter fractions $f_{\mathrm{DM}}(< R_{\mathrm{eff}}) < 0.5$, typically found in massive galaxies, have often been interpreted as evidence for the presence of a central dark matter core \citep{genzel20, nestor23, Danhaive_2025b}.
These claims have motivated theoretical works to explore various mechanisms for the formation of such cores at $z > 1$, including the impact of dry mergers combined with AGN-driven feedback \citep{Dekel_2021}, or the infall of baryonic clumps \citep{Ogiya_2022}. 
However, these observational studies rely on several assumptions, such as warm gas accurately tracing circular motions and on gas fractions and distributions inferred from scaling relations (see Section~\ref{sec5:intro}). 
Moreover, it is worth stressing that nearby massive galaxies typically have low dark matter fractions but cuspy halos \citep[e.g.][]{mancerapina22b}.

The analysis presented in this paper employs data of arguably better quality than that of most previous works, as we have simultaneously available \textit{JWST} observations, spatially resolved cold gas tracers and kinematics. 
Despite having full control of the dominant baryonic components, we have shown that it is not possible to robustly constrain the dark matter content in our galaxies. 
Our fitted virial masses have large errors, and we cannot reliably constrain the halo concentrations even in the simplest case of a 2-parameter fit (Section \ref{sec5:dm}).
We are therefore forced to conclude that previous claims of observational constraints on the inner shape of the dark matter halos at $z > 1$ should be treated with extreme caution.

\begin{figure}
    \centering
    \includegraphics[width=\columnwidth]{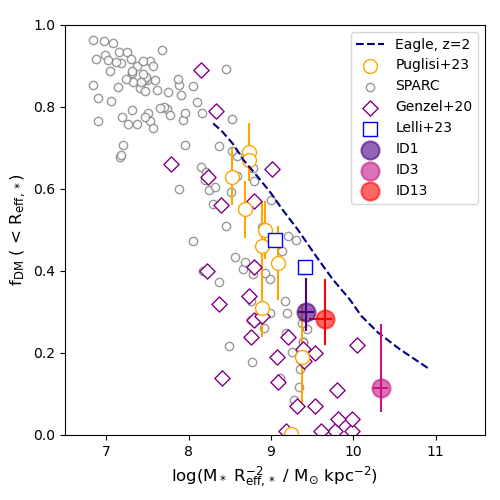}
    \caption[Dark matter fractions vs. stellar mass surface density.]{Dark matter fraction within the stellar disc effective radius versus the stellar mass surface density. We show our three ALPAKA galaxies (filled circles), the KURVS-CDFS galaxies at z $\sim$ 1.5 \citep[open yellow circles, ][]{puglisi23}, the IRAM-NOEMA/RC41 at z $\sim 0.67 - 2.45$ \citep[open purple diamonds, ][]{genzel20}, two ALMA/CO galaxies at z = 2.24 and z=1.47 \citep[open blue squares, ][]{lelli23} and the SPARC galaxies in the local Universe \citep[open gray squares, ][]{lelli16}, according to the legend. 
    We also show the trend for simulated galaxies in EAGLE \citep{crain15}.}
    \label{fig:fdm}
\end{figure}

With the aim of comparing our results and uncertainties with the literature, we have computed $f_{\mathrm{DM}}(< R_{\mathrm{eff}})$ for our galaxies using our fiducial mass models.
In Figure~\ref{fig:fdm}, we show these dark matter fractions within the stellar disc effective radius as a function of stellar mass surface density. 
The dark matter fraction is defined as the mass of dark matter divided by the total mass (baryonic plus dark matter) enclosed within the effective radius of the stellar disc. 
The stellar mass surface density is computed as the total stellar mass divided by the square of the stellar disc effective radius.

We compare our values with the KURVS-CDFS sample from \citet{puglisi23}, the IRAM-NOEMA/RC41 sample from \citet{genzel20}, and the two galaxies from \citet{lelli23}. For reference, we also include local spiral galaxies from the SPARC survey, which contains 175 nearby disc galaxies with high-quality rotation curves traced primarily by H\,I (and H$\alpha$ in the inner regions of massive systems) and 3.6 $\mu$m photometry \citep{lelli16}. 
Finally, we include the trend for simulated galaxies in EAGLE at $z = 2$ \citep{crain15}.
We find that the ALPAKA galaxies, all with dark matter fractions of about 0.2 within the disc effective radius, are broadly consistent with values reported both for local galaxies and, importantly, for systems at $z = 1$–$3$. 
However, our interpretation differs substantially from previous studies. 
Based on our fiducial models of the rotation curve decomposition, these galaxies are well described by standard NFW halos.
In fact, if anything, they may suggest dark matter halos that are cuspier than the NFW profile (Section \ref{sec:alternatives}).

\section{Conclusions}\label{sec5:conclusions}

In this work, we followed up on three cosmic noon ALPAKA galaxies (ID1, ID3 and ID13) with both highly resolved  CO/[CI] emission from ALMA, covering the cold gas, and rest-frame near infrared \textit{JWST} data, covering the bulk of the stellar populations. 
We used the rotation curves obtained with \texttt{$^{\mathrm{3D}}$BAROLO} in \citet{rizzo23}, corrected for pressure support and modelled the mass contribution of the baryonic and dark matter components in the galaxies.
We assumed that the baryons are distributed in an exponential gas disc, an exponential stellar disc and a spherical bulge, and that the dark matter is distributed in an NFW halo.

We reported two models: a fiducial model in which we leave the total mass of the three main components free (gas, stars and dark matter) and a model with the baryonic components fixed to typical values from the literature, considering the SED fits for the stellar mass and the conversion factor $\alpha_{\mathrm{CO/[CI]}}$ and CO line-luminosity ratios for the gas. 
In the fiducial model, we reproduce the rotation curves and obtain dynamical masses for the gas, stars and dark matter. In the fixed baryonic parameters fit, we attempt to model the dark matter parameters.
Here we summarise our main findings:

\begin{itemize}

    \item The fiducial mass models reproduce the overall shape of the rotation curves of our target galaxies, although they systematically underestimate the circular velocities in the inner regions. In these fiducial models, we found that the three galaxies have stellar masses of $\log(M_{\star}/M_{\odot}) = 10.6 \pm 0.1$, $10.9 \pm 0.1$, and $10.7^{+0.1}_{-0.2}$, and dark matter halo masses of $\log(M_{200}/M_{\odot}) = 12.0^{+0.6}_{-0.4}$, $13.5^{+1.7}_{-1.3}$, and $12.5^{+1.0}_{-0.7}$, respectively.

    \item In the fiducial models, we chose to maximise the contribution of the bulges to the rotation curves. 
    This was motivated by the fact that the innermost rotation velocities of our sample require the presence of a massive central component.
    The maximisation of the bulge component was obtained by i) fixing its intensity in the fit of the stellar surface brightness, ii) assuming a factor of 1.4 difference between the mass-to-light ratios ($\Upsilon$) of bulges and discs.
    We investigated possible reasons for ii) and alternatives to the presence of massive bulges.
    We found that both the presence of different stellar populations in the bulge and the disc and the presence of dust attenuation are viable possibilities to explain the different $\Upsilon$ between bulges and discs. 
    These could be further investigated using spatially resolved SED fits with the novel \textit{JWST} data. 
    We also found that SMBHs can contribute significantly to the inner velocities only if the black holes are as massive as recent high-$z$ observations suggest; however, they may not be sufficient to reproduce the high values of the inner rotation velocities on their own. 
    Finally, we found that metallicity gradients in the gas do not significantly improve the fits by driving an $\alpha_{\mathrm{CO}}$ gradient.   
    We consider it probable that more than one of the above phenomena could be at play simultaneously. At the same time, we cannot exclude the possibility that the observed deficit also reflects variations in the inner dark matter distribution, such as concentrations higher than expected in CDM or deviations from the NFW profile due to adiabatic contraction.

    \item We attempted to model the dark matter halos with free parameters by fixing the parameters of the baryonic components to values from the literature. 
    These models are generally worse than our fiducial model in reproducing the rotation curves. 
    ID1 and ID3 become dark matter dominated and both show an issue of tending toward unexpectedly high values for the halo concentrations. 
    The fit of ID3 also returns an extremely large virial mass of $\log(M_{200}/M_{\odot}) = 14.0^{+0.5}_{-0.4}$.
    ID13 continues to be baryon dominated; however, given the estimated gas and stellar masses, it requires a dark matter halo with an unphysically small mass of $\log(M_{200}/M_{\odot}) = 10.4^{+1.1}_{-0.9}$. 
    Overall, this experiment strongly suggests that more precise constraints on the $\alpha_{\mathrm{CO/[CI]}}$ and the CO line-luminosity ratios are needed to constrain the gas masses, which is a necessary condition for inferring precise dark matter halo properties.

    \item The inner dark matter fractions for our galaxies $f_{\mathrm{DM}}(<R_{\mathrm{eff}})\approx0.2$, are comparable to those derived for both local massive discs and $z >1$ star-forming galaxies. However, our interpretation differs from previous studies that have associated such low values with cored dark matter profiles. We find that standard NFW halos can reproduce the observations without invoking cores. 
    If anything, our analysis suggests cuspier halo profiles than the standard NFW.
\end{itemize}

We have carried out a detailed description of the mass budget of galaxies at cosmic noon thanks to the presence of both well resolved cold gas observations with ALMA and rest-frame near-infrared imaging with \textit{JWST}. These promising results are only a glimpse of what can be done with state-of-the-art facilities and much of this analysis can be extended to the remainder of the ALPAKA sample. Considering that there are ALPAKA galaxies with higher gas resolution, once \textit{JWST} data becomes available, we can hope to obtain stronger constraints on their dark matter halo masses and shapes. In addition, forthcoming data for the CONDOR sample will provide robust measurements of both stellar and gas masses, supported by extensive ancillary observations from \textit{JWST}, VLA, and ALMA. This expanded dataset will enable us to further refine our understanding of the mass distributions of cosmic-noon galaxies. Gathering statistics across these samples will also help us validate theoretical predictions regarding the behaviour of high-$z$ dark matter halos and the connection between discs and their host halos.

\section*{Acknowledgements}
We thank Enrico Di Teodoro for valuable discussions. 
FRO and FF acknowledges support from the Dutch Research Council (NWO) through the Klein-1 Grant code OCEN2.KLEIN.088. FR acknowledges support from the Dutch Research Council (NWO) through the Veni grant VI.Veni.222.146. 
This work has received funding from the European Research Council (ERC) under the Horizon Europe research and innovation programme (Acronym: FLOWS, Grant number: 101096087). ALMA is a partnership of ESO (representing its member states), NSF (USA) and NINS (Japan), together with NRC (Canada), NSC and ASIAA (Taiwan), and KASI (Republic of Korea), in cooperation with the Republic of Chile. The Joint ALMA Observatory is operated by ESO, AUI/NRAO and NAOJ. This work is based on observations made with the NASA/ESA/CSA James Webb Space Telescope. The data were obtained from the Mikulski Archive for Space Telescopes at the Space Telescope Science Institute, which is operated by the Association of Universities for Research in Astronomy, Inc., under NASA contract NAS 5-03127 for JWST. The JWST data underlying this article are publicly available and were accessed from the DAWN JWST Archive (DJA). DJA is an initiative of the Cosmic Dawn Center (DAWN), which is funded by the Danish National Research Foundation under grant DNRF140.

We acknowledge usage of the Python programming language \citep{vanrossum09}, Astropy \citep{astropycollab22}, Matplotlib \citep{hunter07}, NumPy \citep{vanderwalt11}, and SciPy \citep{virtanen20}.

\section*{Data Availability}
The JWST observations used in this paper are publicly available and are associated with program \#1180, \#1727, \#1837 and
can be obtained from
\url{https://dx.doi.org/10.17909/6qq5-w568}. The ALMA data are publicly available and can be retrieved through the ALMA Archive website. The ALMA data are associated with the following programs: ADS/JAO.ALMA\#2017.1.01659.S; ADS/JAO.ALMA\#2017.1.00471.S; ADS/JAO.ALMA\#2016.1.01426.S; ADS/JAO.ALMA\#2017.1.00413.S.

\bibliographystyle{mnras}
\bibliography{references}

\begin{appendix}

\section{Gas Surface Brightness Best-fit Posteriors}\label{sec5:dysb_corner}

\begin{figure*}
    \centering
    \includegraphics[width=0.9\textwidth]{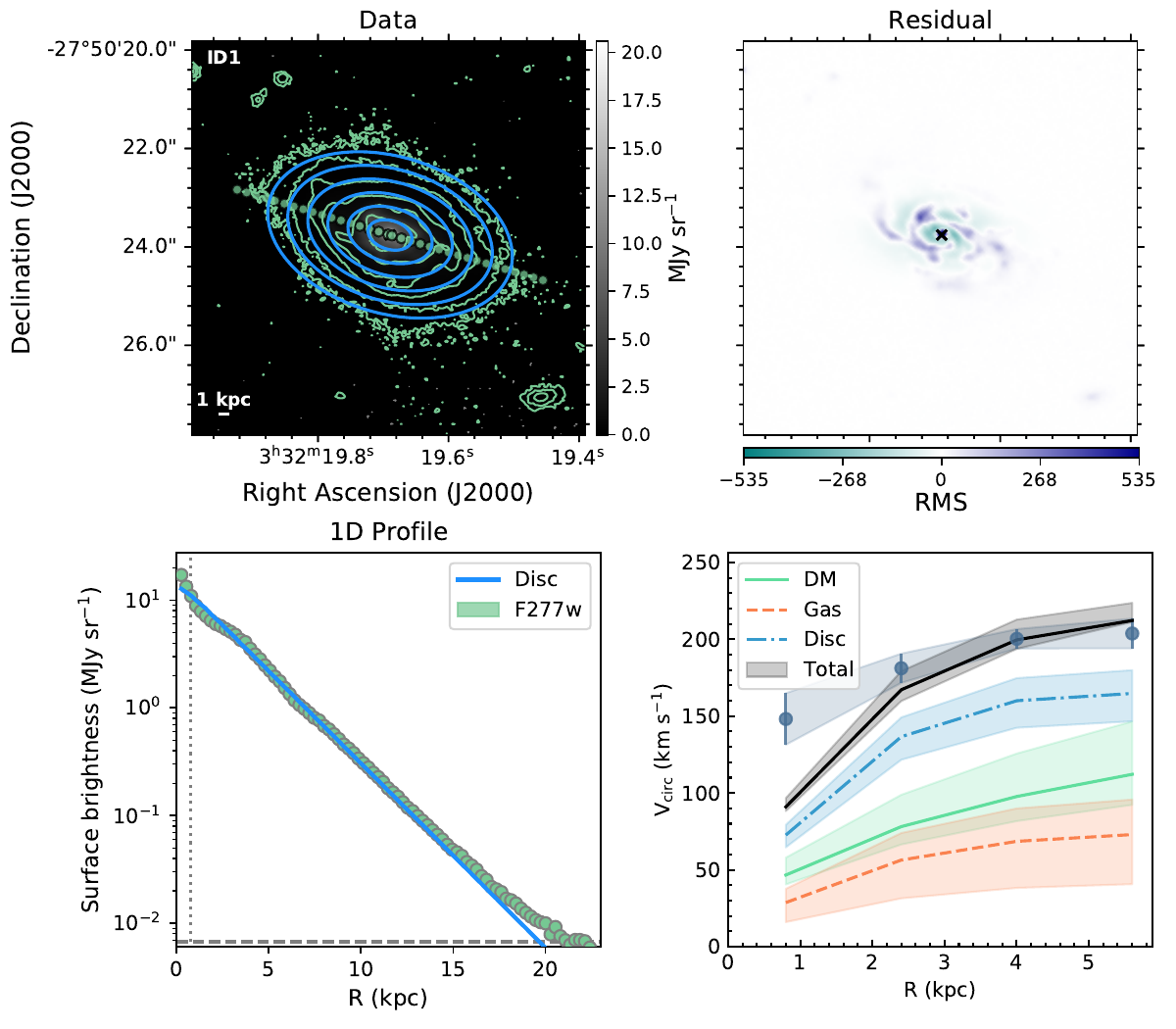}
    \caption[Single exponential model of the gas surface brightness of ID1 and its best-fit mass model.]{Results of the surface brightness fit of ID1 with one stellar disc component. Top left panel: the \textit{JWST} image in grayscale is overlayed with contours for the data (teal) and for the best-fit model (blue), following levels of 3$\sigma$, 9$\sigma$, 27$\sigma$, etc., where $\sigma$ is the RMS noise in the data. Top right panel: noise-normalised residual map defined by (Data - Model) / RMS. Bottom left panel: 1D representation of the projected average surface brightness over rings along the major axis of the galaxy. The teal band represents the data and the blue solid line represents the total best-fit model which is composed by a single the stellar disc. Bottom right panel: the best-fit mass model considering this surface brightness fit.}
    \label{fig:nobulge}
\end{figure*}

\begin{figure*}
    \centering
    \includegraphics[width=0.99\linewidth]{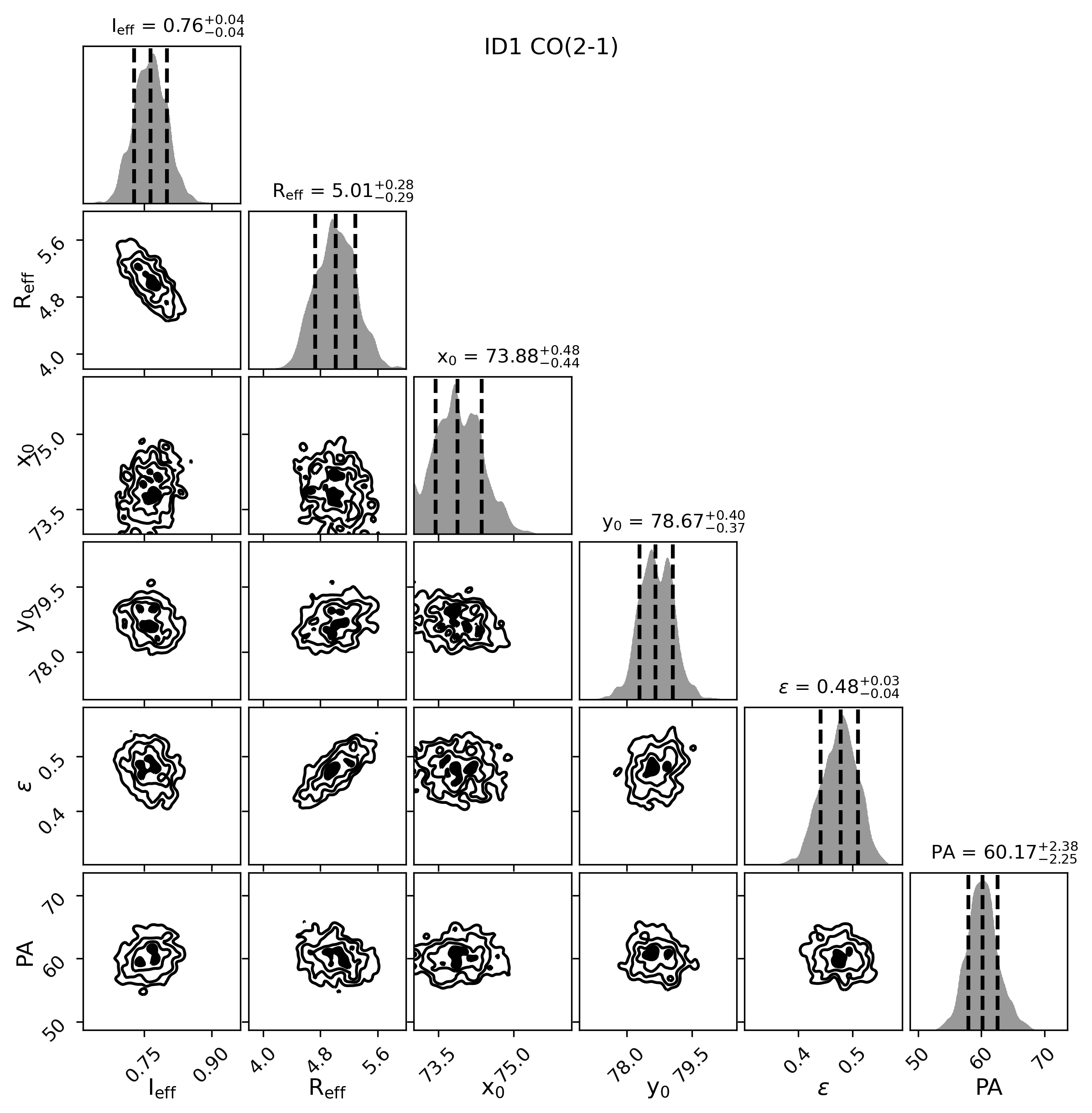}
    \caption[Posterior distribution of the gas surface brightness of ID1.]{Cornerplot of the best-fit parameters for the gas distribution of ID1 considering a single exponential disc component, the best-fit model is shown in Figure~\ref{fig:sb_id1}. We show the intensity of the CO(2-1) 0th-moment map at the effective radius in mJy/beam$\kms$ (I$_{\mathrm{eff}}$), the effective radius in kpc (R$_{eff}$), centre ($x_0$ and $y_0$ in pixels), disc ellipticity ($\epsilon$) and position angle in degrees (PA).}
    \label{fig:dysb_id1_corner}
\end{figure*}

\begin{figure*}
    \centering
    \includegraphics[width=0.99\linewidth]{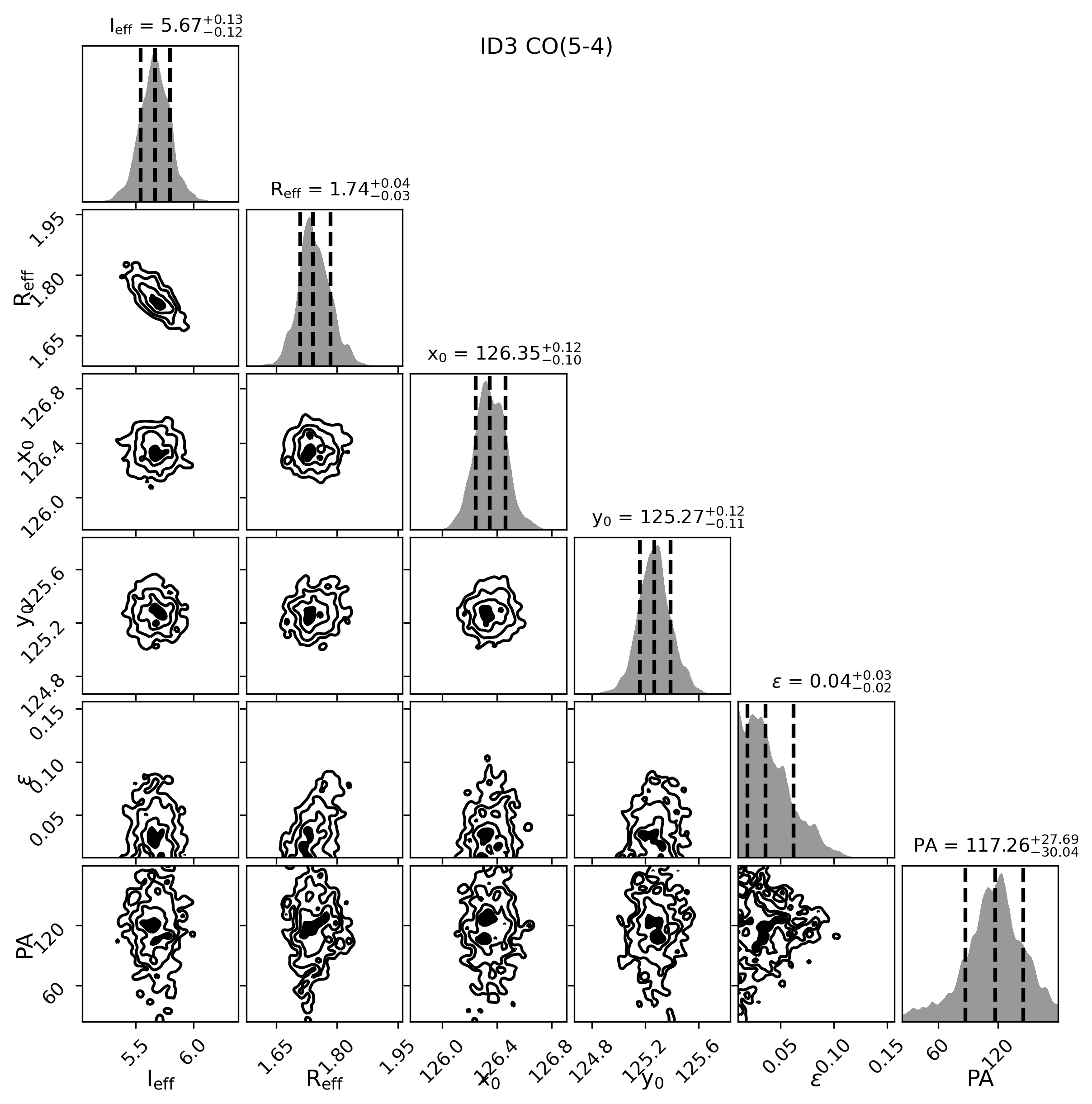}
    \caption[Posterior distribution of the gas surface brightness of ID3.]{Same as Figure~\ref{fig:dysb_id1_corner}, but for the best-fit model of CO(5-4) emission of ID3 shown in Figure~\ref{fig:sb_id3}.}
    \label{fig:dysb_id3_corner}
\end{figure*}

\begin{figure*}
    \centering
    \includegraphics[width=0.99\linewidth]{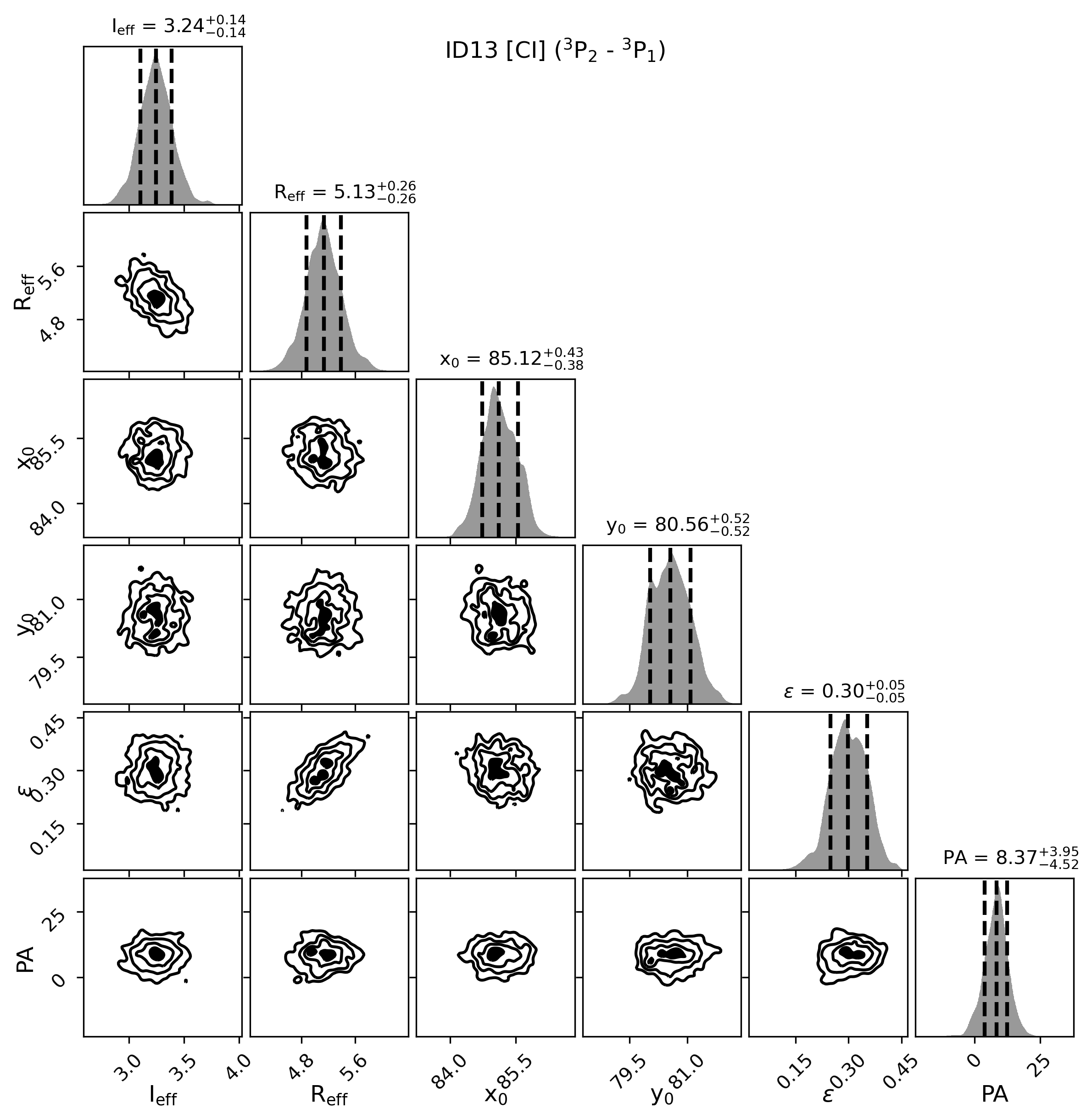}
    \caption[Posterior distribution of the gas surface brightness of ID13.]{Same as Figure~\ref{fig:dysb_id1_corner}, but for the best-fit model of [CI] ($^{3}$P$_{2}$ - $^{3}$P$_{1}$) emission of ID13 shown in Figure~\ref{fig:sb_id13}.}
    \label{fig:dysb_id13_corner}
\end{figure*}

In the first attempt to model the two-dimensional surface brightness of the stellar component, we considered only an exponential disc. Figure~\ref{fig:nobulge} shows a representative example for ID1. The surface brightness profile is reasonably well reproduced by this single-component model, although it slightly underestimates the observed brightness in the innermost region, within the spatial resolution limit of the data (vertical gray line). However, when this stellar distribution is adopted, the corresponding dynamical model fails to reproduce the observed rotation curve (bottom-left panel in Figure~\ref{fig:nobulge}). We show the example of ID1; however, we note that this also occurs for ID3 and ID13. 
For this reason, for the fiducial models shown in Section~\ref{sec5:fid}, we chose to obtain 2D models of the galaxies in a way that maximises the brightness of the bulges in the three galaxies. In Figures~\ref{fig:dysb_id1_corner} - ~\ref{fig:dysb_id13_corner}, we show the posterior of the best-fit parameters of the 2D surface brightness fit model of the gas emission of ID1, ID3 and ID13.

\section{Asymmetric Drift Correction}\label{sec5:adc}
\begin{figure*}
    \centering
    \includegraphics[width=0.92\textwidth]{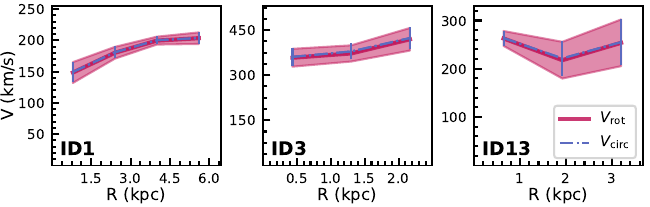}
    \caption[Asymmetric drift correction of selected ALPAKA galaxies.]{Rotation curves corrected for pressure support. We show the rotation velocity obtained with \texttt{$^{\mathrm{3D}}$BAROLO} ($V_{\mathrm{rot}}$) in magenta and the circular speed ($V_{\mathrm{circ}}$) in violet obtained after applying the asymmetric drift correction.}
    \label{fig5:adc}
\end{figure*}

In Figure~\ref{fig5:adc}, we show the result of the correction of the rotation curve for pressure support by following the asymmetric drift correction explained in Section~\ref{met:adc}.

\section{Fiducial mass model cornerplots}\label{sec5:dydy}
\begin{figure*}
    \centering
    \includegraphics[width=0.99\textwidth]{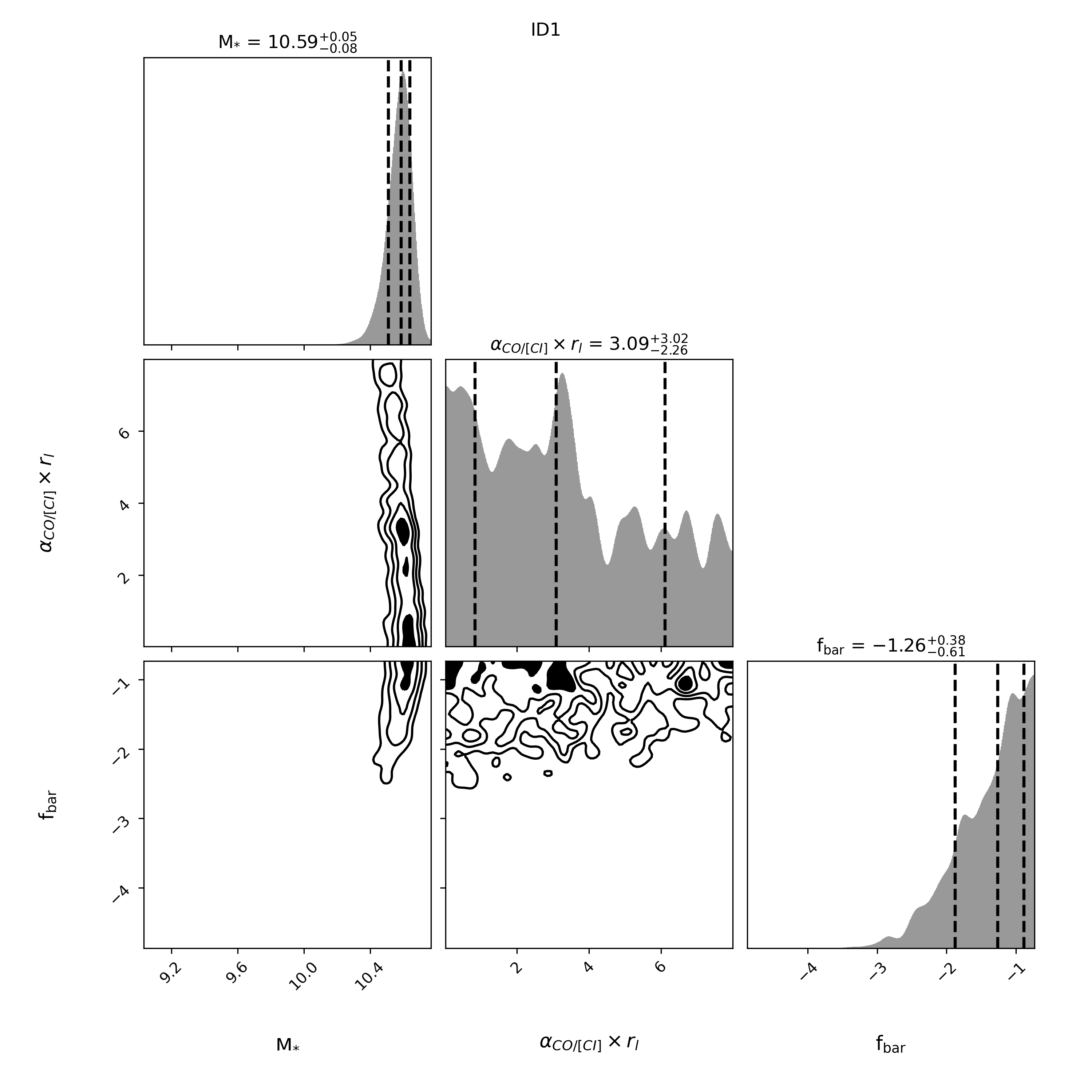}
    \caption[Posterior distribution of the best-fit mass model of ID1.]{The posteriors of the best-fit parameters for the fiducial mass models of ID1, shown in Figure~\ref{fig:fid}. We show the total stellar mass in log space (M$_{*}$); the gas normalisation following Equation~\ref{eq:alpha} that consists of the $\alpha_{\mathrm{CO}}$ and the line ratio $r_l$; the baryon fraction in log space (f$_{\mathrm{bar}}$).}
    \label{fig:dydy_ID1_corner}
\end{figure*}

\begin{figure*}
    \centering
    \includegraphics[width=0.99\textwidth]{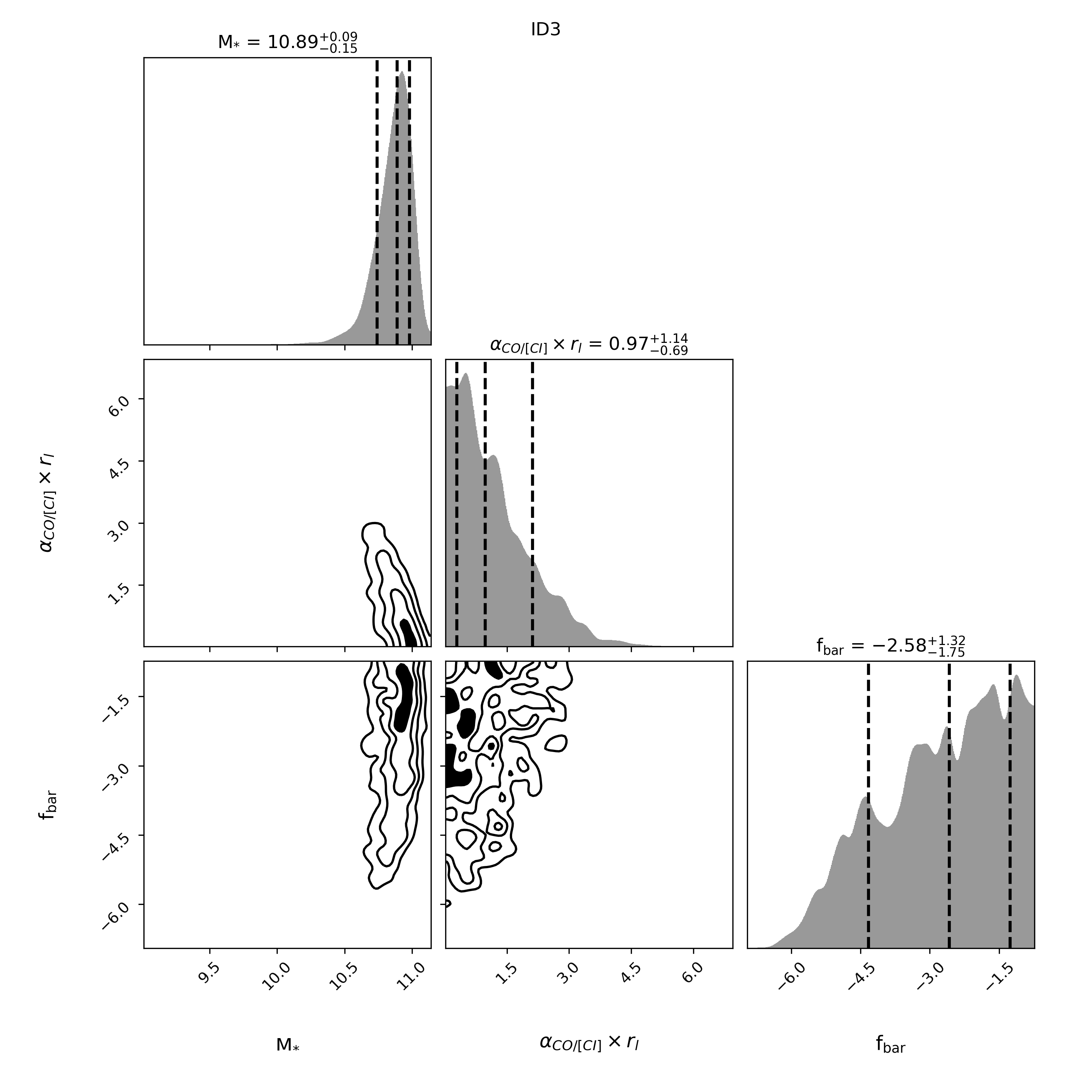}
    \caption[Posterior distribution of the best-fit mass model of ID3.]{Same as Figure~\ref{fig:dydy_ID1_corner}, but for ID3.}
    \label{fig:dydy_ID3_corner}
\end{figure*}

\begin{figure*}
    \centering
    \includegraphics[width=0.99\textwidth]{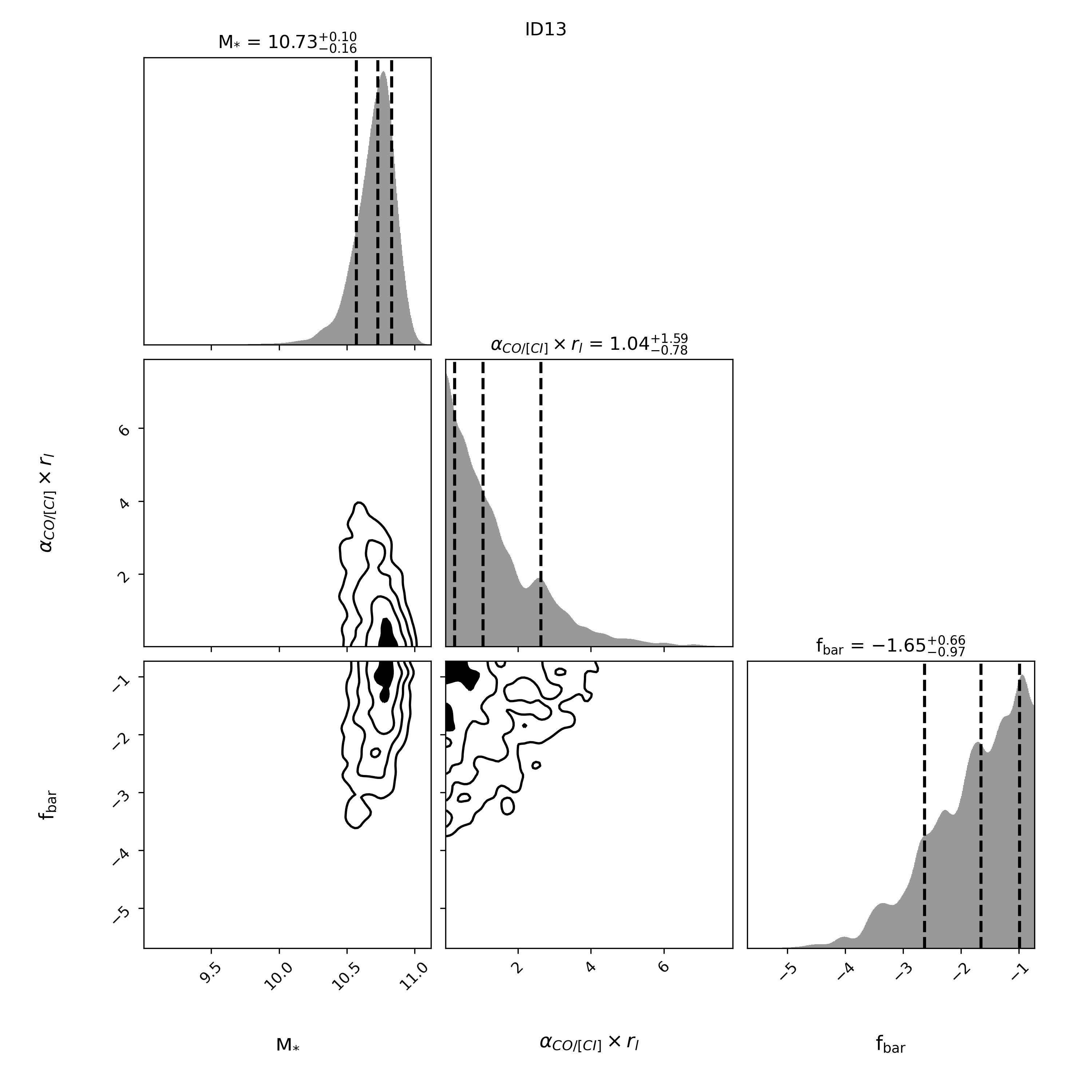}
    \caption[Posterior distribution of the best-fit mass model of ID13.]{Same as Figure~\ref{fig:dydy_ID1_corner}, but for ID13.}
    \label{fig:dydy_ID13_corner}
\end{figure*}

\begin{figure*}
    \centering
    \includegraphics[width=0.99\textwidth]{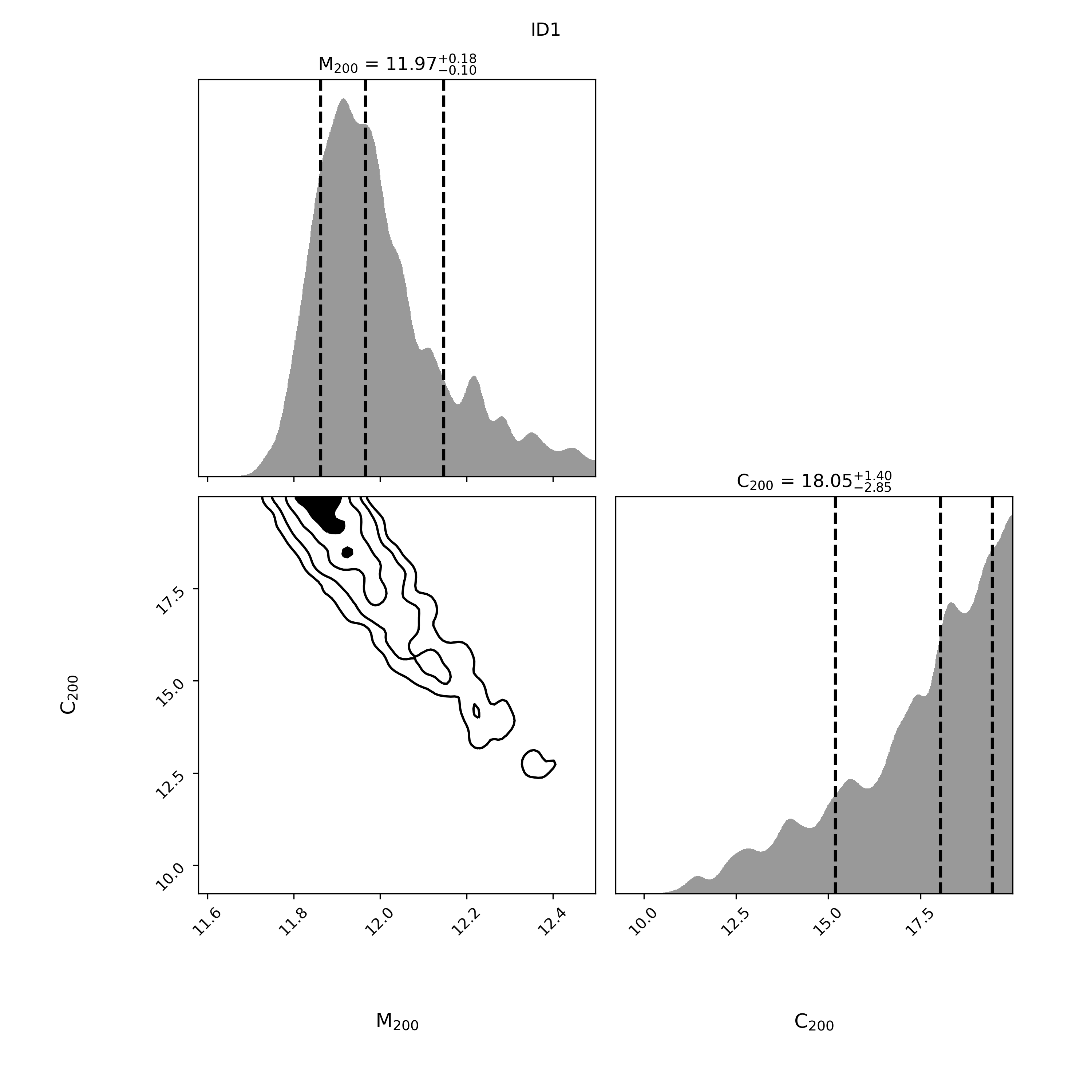}
    \caption[Posterior distribution of the best-fit mass model with fixed baryonic parameters of ID1.]{The posteriors of the best-fit parameters for mass models with the baryonic component parameters fixed shown in Figure~\ref{fig:alt}. We show the total dark matter NFW halo mass (M$_{200}$) and concentration (c$_{200}$).}
    \label{fig:dydy_ID1_corner_alt}
\end{figure*}

\begin{figure*}
    \centering
    \includegraphics[width=0.99\textwidth]{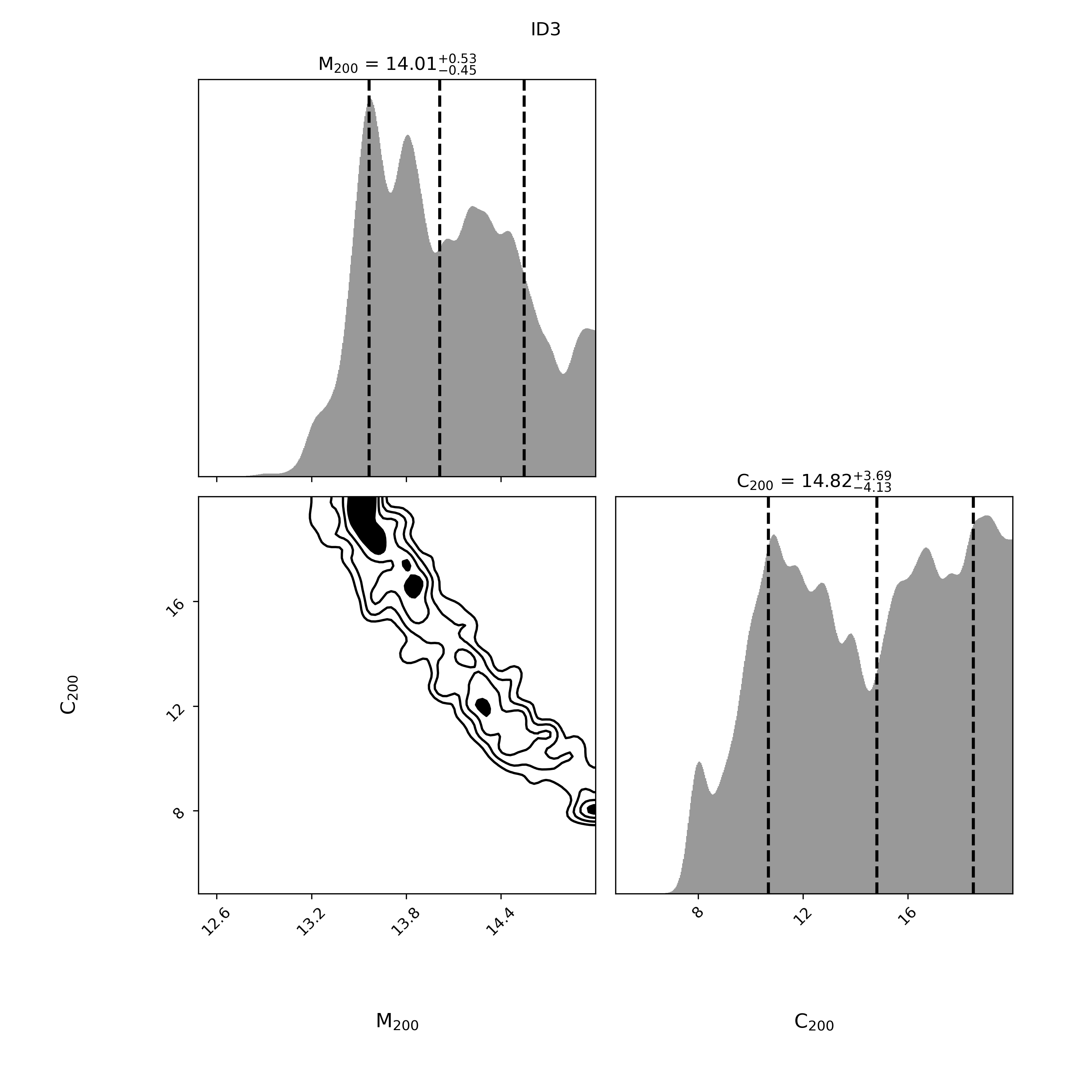}
    \caption[Posterior distribution of the best-fit mass model with fixed baryonic parameters of ID3.]{Same as Figure~\ref{fig:dydy_ID1_corner_alt}, but for ID3.}
    \label{fig:dydy_ID3_corner_alt}
\end{figure*}

\begin{figure*}
    \centering
    \includegraphics[width=0.99\textwidth]{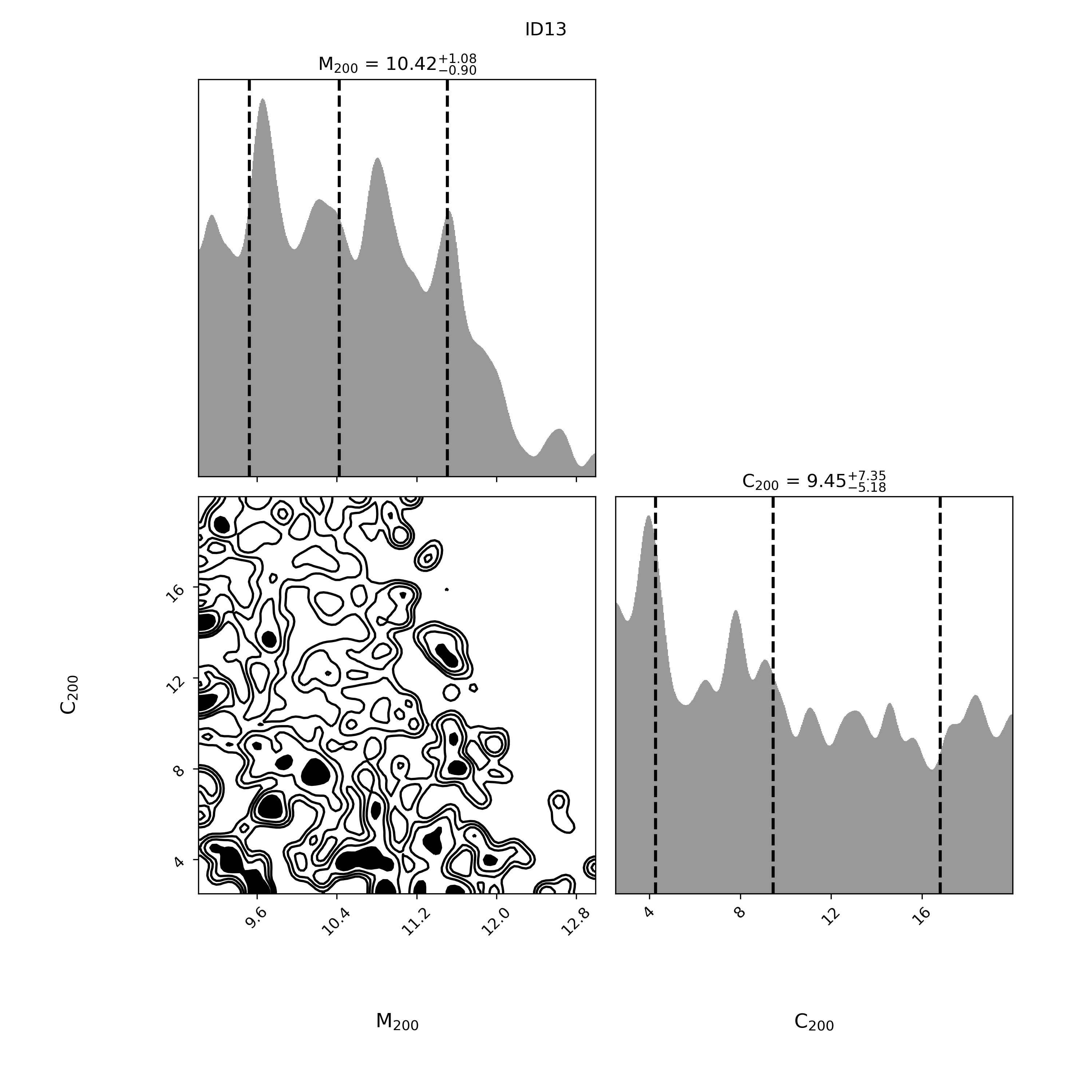}
    \caption[Posterior distribution of the best-fit mass model with fixed baryonic parameters of ID13.]{Same as Figure~\ref{fig:dydy_ID1_corner_alt}, but for ID13.}
    \label{fig:dydy_ID13_corner_alt}
\end{figure*}

In Figures~\ref{fig:dydy_ID1_corner} --\ref{fig:dydy_ID13_corner}, we show the posteriors of the best-fit parameters for the fiducial rotation curve decomposition shown in Section~\ref{sec5:fid}. In Figures~\ref{fig:dydy_ID1_corner_alt} -- \ref{fig:dydy_ID13_corner_alt}, we show the posteriors for the mass models with the baryonic component parameters kept fixed where we try constraining the properties of the dark matter halo in Section~\ref{sec5:dm}.

\end{appendix}

\bsp
\label{lastpage}

\end{document}